\documentclass[twoside]{IEEEtran} 
\pdfoutput=1 

\usepackage[intlimits,cmex10]{amsmath}
\usepackage{xspace}
\usepackage[nosort]{cite}        
\usepackage{url} 
\usepackage{bbm}
\usepackage{graphicx}
\usepackage{subfigure}         
\usepackage{paralist}
\usepackage{fancyref}
\usepackage[stretch=16,shrink=16,step=4]{microtype}
\usepackage{pdfsync}
\usepackage{color}
\newlength{\figwidth}
\makeatletter
\def\@IEEEinterspaceratioM{0.265}
\def\@IEEEinterspaceMINratioM{0.1651}
\def\@IEEEinterspaceMAXratioM{0.38}

\def\@IEEEinterspaceratioB{0.31}
\def\@IEEEinterspaceMINratioB{0.19}
\def\@IEEEinterspaceMAXratioB{0.38}
\@IEEEtunefonts
\makeatother

\usepackage{amssymb}
\usepackage{amsfonts}
\usepackage{mathrsfs}
\usepackage{xspace}
\usepackage{bm}
\usepackage{upgreek}

\newcommand{\safemath}[2]{\newcommand{#1}{\ensuremath{#2}\xspace}}

\safemath{\bma}{\mathbf{a}}
\safemath{\bmb}{\mathbf{b}}
\safemath{\bmc}{\mathbf{c}}
\safemath{\bmd}{\mathbf{d}}
\safemath{\bme}{\mathbf{e}}
\safemath{\bmf}{\mathbf{f}}
\safemath{\bmg}{\mathbf{g}}
\safemath{\bmh}{\mathbf{h}}
\safemath{\bmi}{\mathbf{i}}
\safemath{\bmj}{\mathbf{j}}
\safemath{\bmk}{\mathbf{k}}
\safemath{\bml}{\mathbf{l}}
\safemath{\bmm}{\mathbf{m}}
\safemath{\bmn}{\mathbf{n}}
\safemath{\bmo}{\mathbf{o}}
\safemath{\bmp}{\mathbf{p}}
\safemath{\bmq}{\mathbf{q}}
\safemath{\bmr}{\mathbf{r}}
\safemath{\bms}{\mathbf{s}}
\safemath{\bmt}{\mathbf{t}}
\safemath{\bmu}{\mathbf{u}}
\safemath{\bmv}{\mathbf{v}}
\safemath{\bmw}{\mathbf{w}}
\safemath{\bmx}{\mathbf{x}}
\safemath{\bmy}{\mathbf{y}}
\safemath{\bmz}{\mathbf{z}}
\safemath{\bmzero}{\mathbf{0}}
\safemath{\bmone}{\mathbf{1}}

\bmdefine{\biad}{a}
\bmdefine{\bibd}{b}
\bmdefine{\bicd}{c}
\bmdefine{\bidd}{d}
\bmdefine{\bied}{e}
\bmdefine{\bifd}{f}
\bmdefine{\bigd}{g}
\bmdefine{\bihd}{h}
\bmdefine{\biid}{i}
\bmdefine{\bijd}{j}
\bmdefine{\bikd}{k}
\bmdefine{\bild}{l}
\bmdefine{\bimd}{m}
\bmdefine{\bind}{n}
\bmdefine{\biod}{o}
\bmdefine{\bipd}{p}
\bmdefine{\biqd}{q}
\bmdefine{\bird}{r}
\bmdefine{\bisd}{s}
\bmdefine{\bitd}{t}
\bmdefine{\biud}{u}
\bmdefine{\bivd}{v}
\bmdefine{\biwd}{w}
\bmdefine{\bixd}{x}
\bmdefine{\biyd}{y}
\bmdefine{\bizd}{z}

\bmdefine{\bixid}{\xi}
\bmdefine{\bilambdad}{\lambda}
\bmdefine{\bimud}{\mu}
\bmdefine{\bithetad}{\theta}
\bmdefine{\biphid}{\phi}

\safemath{\bmia}{\biad}
\safemath{\bmib}{\bibd}
\safemath{\bmic}{\bicd}
\safemath{\bmid}{\bidd}
\safemath{\bmie}{\bied}
\safemath{\bmif}{\bifd}
\safemath{\bmig}{\bigd}
\safemath{\bmih}{\bihd}
\safemath{\bmii}{\biid}
\safemath{\bmij}{\bijd}
\safemath{\bmik}{\bikd}
\safemath{\bmil}{\bild}
\safemath{\bmim}{\bimd}
\safemath{\bmin}{\bind}
\safemath{\bmio}{\biod}
\safemath{\bmip}{\bipd}
\safemath{\bmiq}{\biqd}
\safemath{\bmir}{\bird}
\safemath{\bmis}{\bisd}
\safemath{\bmit}{\bitd}
\safemath{\bmiu}{\biud}
\safemath{\bmiv}{\bivd}
\safemath{\bmiw}{\biwd}
\safemath{\bmix}{\bixd}
\safemath{\bmiy}{\biyd}
\safemath{\bmiz}{\bizd}

\safemath{\bmxi}{\bixid}
\safemath{\bmlambda}{\bilambdad}
\safemath{\bmmu}{\bimud}
\safemath{\bmtheta}{\bithetad}
\safemath{\bmphi}{\biphid}

\safemath{\bA}{\mathbf{A}}
\safemath{\bB}{\mathbf{B}}
\safemath{\bC}{\mathbf{C}}
\safemath{\bD}{\mathbf{D}}
\safemath{\bE}{\mathbf{E}}
\safemath{\bF}{\mathbf{F}}
\safemath{\bG}{\mathbf{G}}
\safemath{\bH}{\mathbf{H}}
\safemath{\bI}{\mathbf{I}}
\safemath{\bJ}{\mathbf{J}}
\safemath{\bK}{\mathbf{K}}
\safemath{\bL}{\mathbf{L}}
\safemath{\bM}{\mathbf{M}}
\safemath{\bN}{\mathbf{N}}
\safemath{\bO}{\mathbf{O}}
\safemath{\bP}{\mathbf{P}}
\safemath{\bQ}{\mathbf{Q}}
\safemath{\bR}{\mathbf{R}}
\safemath{\bS}{\mathbf{S}}
\safemath{\bT}{\mathbf{T}}
\safemath{\bU}{\mathbf{U}}
\safemath{\bV}{\mathbf{V}}
\safemath{\bW}{\mathbf{W}}
\safemath{\bX}{\mathbf{X}}
\safemath{\bY}{\mathbf{Y}}
\safemath{\bZ}{\mathbf{Z}}

\safemath{\bZero}{\mathbf{0}}
\safemath{\bOne}{\mathbf{1}}
\safemath{\bDelta}{\mathbf{\Delta}}
\safemath{\bLambda}{\mathbf{\UpLambda}}
\safemath{\bPhi}{\mathbf{\Upphi}}
\safemath{\bSigma}{\mathbf{\Upsigma}}
\safemath{\bOmega}{\mathbf{\Upomega}}
\safemath{\bTheta}{\mathbf{\Uptheta}}

\bmdefine{\biAd}{A}
\bmdefine{\biBd}{B}
\bmdefine{\biCd}{C}
\bmdefine{\biDd}{D}
\bmdefine{\biEd}{E}
\bmdefine{\biFd}{F}
\bmdefine{\biGd}{G}
\bmdefine{\biHd}{H}
\bmdefine{\biId}{I}
\bmdefine{\biJd}{J}
\bmdefine{\biKd}{K}
\bmdefine{\biLd}{L}
\bmdefine{\biMd}{M}
\bmdefine{\biOd}{N}
\bmdefine{\biPd}{O}
\bmdefine{\biQd}{P}
\bmdefine{\biRd}{R}
\bmdefine{\biSd}{S}
\bmdefine{\biTd}{T}
\bmdefine{\biUd}{U}
\bmdefine{\biVd}{V}
\bmdefine{\biWd}{W}
\bmdefine{\biXd}{X}
\bmdefine{\biYd}{Y}
\bmdefine{\biZd}{Z}

\bmdefine{\biDelta}{\Delta}
\bmdefine{\biLambda}{\Lambda}
\bmdefine{\biPhi}{\Phi}
\bmdefine{\biSigma}{\Sigma}
\bmdefine{\biOmega}{\Omega}
\bmdefine{\biTheta}{\Theta}

\safemath{\bimA}{\biAd}
\safemath{\bimB}{\biBd}
\safemath{\bimC}{\biCd}
\safemath{\bimD}{\biDd}
\safemath{\bimE}{\biEd}
\safemath{\bimF}{\biFd}
\safemath{\bimG}{\biGd}
\safemath{\bimH}{\biHd}
\safemath{\bimI}{\biId}
\safemath{\bimJ}{\biJd}
\safemath{\bimK}{\biKd}
\safemath{\bimL}{\biLd}
\safemath{\bimM}{\biMd}
\safemath{\bimN}{\biNd}
\safemath{\bimO}{\biOd}
\safemath{\bimP}{\biPd}
\safemath{\bimQ}{\biQd}
\safemath{\bimR}{\biRd}
\safemath{\bimS}{\biSd}
\safemath{\bimT}{\biTd}
\safemath{\bimU}{\biUd}
\safemath{\bimV}{\biVd}
\safemath{\bimW}{\biWd}
\safemath{\bimX}{\biXd}
\safemath{\bimY}{\biYd}
\safemath{\bimZ}{\biZd}

\safemath{\bimDelta}{\biDelta}
\safemath{\bimLambda}{\biLambda}
\safemath{\bimPhi}{\biPhi}
\safemath{\bimSigma}{\biSigma}
\safemath{\bimOmega}{\biOmega}
\safemath{\bimTheta}{\biTheta}

\safemath{\setA}{\mathcal{A}}
\safemath{\setB}{\mathcal{B}}
\safemath{\setC}{\mathcal{C}}
\safemath{\setD}{\mathcal{D}}
\safemath{\setE}{\mathcal{E}}
\safemath{\setF}{\mathcal{F}}
\safemath{\setG}{\mathcal{G}}
\safemath{\setH}{\mathcal{H}}
\safemath{\setI}{\mathcal{I}}
\safemath{\setJ}{\mathcal{J}}
\safemath{\setK}{\mathcal{K}}
\safemath{\setL}{\mathcal{L}}
\safemath{\setM}{\mathcal{M}}
\safemath{\setN}{\mathcal{N}}
\safemath{\setO}{\mathcal{O}}
\safemath{\setP}{\mathcal{P}}
\safemath{\setQ}{\mathcal{Q}}
\safemath{\setR}{\mathcal{R}}
\safemath{\setS}{\mathcal{S}}
\safemath{\setT}{\mathcal{T}}
\safemath{\setU}{\mathcal{U}}
\safemath{\setV}{\mathcal{V}}
\safemath{\setW}{\mathcal{W}}
\safemath{\setX}{\mathcal{X}}
\safemath{\setY}{\mathcal{Y}}
\safemath{\setZ}{\mathcal{Z}}
\safemath{\emptySet}{\varnothing}

\safemath{\colA}{\mathscr{A}}
\safemath{\colB}{\mathscr{B}}
\safemath{\colC}{\mathscr{C}}
\safemath{\colD}{\mathscr{D}}
\safemath{\colE}{\mathscr{E}}
\safemath{\colF}{\mathscr{F}}
\safemath{\colG}{\mathscr{G}}
\safemath{\colH}{\mathscr{H}}
\safemath{\colI}{\mathscr{I}}
\safemath{\colJ}{\mathscr{J}}
\safemath{\colK}{\mathscr{K}}
\safemath{\colL}{\mathscr{L}}
\safemath{\colM}{\mathscr{M}}
\safemath{\colN}{\mathscr{N}}
\safemath{\colO}{\mathscr{O}}
\safemath{\colP}{\mathscr{P}}
\safemath{\colQ}{\mathscr{Q}}
\safemath{\colR}{\mathscr{R}}
\safemath{\colS}{\mathscr{S}}
\safemath{\colT}{\mathscr{T}}
\safemath{\colU}{\mathscr{U}}
\safemath{\colV}{\mathscr{V}}
\safemath{\colW}{\mathscr{W}}
\safemath{\colX}{\mathscr{X}}
\safemath{\colY}{\mathscr{Y}}
\safemath{\colZ}{\mathscr{Z}}

\safemath{\opA}{\mathbb{A}}
\safemath{\opB}{\mathbb{B}}
\safemath{\opC}{\mathbb{C}}
\safemath{\opD}{\mathbb{D}}
\safemath{\opE}{\mathbb{E}}
\safemath{\opF}{\mathbb{F}}
\safemath{\opG}{\mathbb{G}}
\safemath{\opH}{\mathbb{H}}
\safemath{\opI}{\mathbb{I}}
\safemath{\opJ}{\mathbb{J}}
\safemath{\opK}{\mathbb{K}}
\safemath{\opL}{\mathbb{L}}
\safemath{\opM}{\mathbb{M}}
\safemath{\opN}{\mathbb{N}}
\safemath{\opO}{\mathbb{O}}
\safemath{\opP}{\mathbb{P}}
\safemath{\opQ}{\mathbb{Q}}
\safemath{\opR}{\mathbb{R}}
\safemath{\opS}{\mathbb{S}}
\safemath{\opT}{\mathbb{T}}
\safemath{\opU}{\mathbb{U}}
\safemath{\opV}{\mathbb{V}}
\safemath{\opW}{\mathbb{W}}
\safemath{\opX}{\mathbb{X}}
\safemath{\opY}{\mathbb{Y}}
\safemath{\opZ}{\mathbb{Z}}
\safemath{\opZero}{\mathbb{O}}
\safemath{\identityop}{\opI}

\safemath{\veca}{\bma}
\safemath{\vecb}{\bmb}
\safemath{\vecc}{\bmc}
\safemath{\vecd}{\bmd}
\safemath{\vece}{\bme}
\safemath{\vecf}{\bmf}
\safemath{\vecg}{\bmg}
\safemath{\vech}{\bmh}
\safemath{\veci}{\bmi}
\safemath{\vecj}{\bmj}
\safemath{\veck}{\bmk}
\safemath{\vecl}{\bml}
\safemath{\vecm}{\bmm}
\safemath{\vecn}{\bmn}
\safemath{\veco}{\bmo}
\safemath{\vecp}{\bmp}
\safemath{\vecq}{\bmq}
\safemath{\vecr}{\bmr}
\safemath{\vecs}{\bms}
\safemath{\vect}{\bmt}
\safemath{\vecu}{\bmu}
\safemath{\vecv}{\bmv}
\safemath{\vecw}{\bmw}
\safemath{\vecx}{\bmx}
\safemath{\vecy}{\bmy}
\safemath{\vecz}{\bmz}

\safemath{\veczero}{\bmzero}
\safemath{\vecone}{\bmone}
\safemath{\vecxi}{\bmxi}
\safemath{\veclambda}{\bmlambda}
\safemath{\vecmu}{\bmmu}
\safemath{\vectheta}{\bmtheta}
\safemath{\vecphi}{\bmphi}

\safemath{\matA}{\bA}
\safemath{\matB}{\bB}
\safemath{\matC}{\bC}
\safemath{\matD}{\bD}
\safemath{\matE}{\bE}
\safemath{\matF}{\bF}
\safemath{\matG}{\bG}
\safemath{\matH}{\bH}
\safemath{\matI}{\bI}
\safemath{\matJ}{\bJ}
\safemath{\matK}{\bK}
\safemath{\matL}{\bL}
\safemath{\matM}{\bM}
\safemath{\matN}{\bN}
\safemath{\matO}{\bO}
\safemath{\matP}{\bP}
\safemath{\matQ}{\bQ}
\safemath{\matR}{\bR}
\safemath{\matS}{\bS}
\safemath{\matT}{\bT}
\safemath{\matU}{\bU}
\safemath{\matV}{\bV}
\safemath{\matW}{\bW}
\safemath{\matX}{\bX}
\safemath{\matY}{\bY}
\safemath{\matZ}{\bZ}
\safemath{\matzero}{\bmzero}

\safemath{\matDelta}{\bDelta}
\safemath{\matLambda}{\bLambda}
\safemath{\matPhi}{\bPhi}
\safemath{\matSigma}{\bSigma}
\safemath{\matOmega}{\bOmega}
\safemath{\matTheta}{\bTheta}

\safemath{\matidentity}{\matI}
\safemath{\matone}{\matO}

\safemath{\rnda}{A}
\safemath{\rndb}{B}
\safemath{\rndc}{C}
\safemath{\rndd}{D}
\safemath{\rnde}{E}
\safemath{\rndf}{F}
\safemath{\rndg}{G}
\safemath{\rndh}{H}
\safemath{\rndi}{I}
\safemath{\rndj}{J}
\safemath{\rndk}{K}
\safemath{\rndl}{L}
\safemath{\rndm}{M}
\safemath{\rndn}{N}
\safemath{\rndo}{O}
\safemath{\rndp}{P}
\safemath{\rndq}{Q}
\safemath{\rndr}{R}
\safemath{\rnds}{S}
\safemath{\rndt}{T}
\safemath{\rndu}{U}
\safemath{\rndv}{V}
\safemath{\rndw}{W}
\safemath{\rndx}{X}
\safemath{\rndy}{Y}
\safemath{\rndz}{Z}

\safemath{\rveca}{\bimA}
\safemath{\rvecb}{\bimB}
\safemath{\rvecc}{\bimC}
\safemath{\rvecd}{\bimD}
\safemath{\rvece}{\bimE}
\safemath{\rvecf}{\bimF}
\safemath{\rvecg}{\bimG}
\safemath{\rvech}{\bimH}
\safemath{\rveci}{\bimI}
\safemath{\rvecj}{\bimJ}
\safemath{\rveck}{\bimK}
\safemath{\rvecl}{\bimL}
\safemath{\rvecm}{\bimM}
\safemath{\rvecn}{\bimN}
\safemath{\rveco}{\bomO}
\safemath{\rvecp}{\bimP}
\safemath{\rvecq}{\bimQ}
\safemath{\rvecr}{\bimR}
\safemath{\rvecs}{\bimS}
\safemath{\rvect}{\bimT}
\safemath{\rvecu}{\bimU}
\safemath{\rvecv}{\bimV}
\safemath{\rvecw}{\bimW}
\safemath{\rvecx}{\bimX}
\safemath{\rvecy}{\bimY}
\safemath{\rvecz}{\bimZ}

\safemath{\rvecxi}{\bmxi}
\safemath{\rveclambda}{\bmlambda}
\safemath{\rvecmu}{\bmmu}
\safemath{\rvectheta}{\bmtheta}
\safemath{\rvecphi}{\bmphi}

\safemath{\rmatA}{\bimA}
\safemath{\rmatB}{\bimB}
\safemath{\rmatC}{\bimC}
\safemath{\rmatD}{\bimD}
\safemath{\rmatE}{\bimE}
\safemath{\rmatF}{\bimF}
\safemath{\rmatG}{\bimG}
\safemath{\rmatH}{\bimH}
\safemath{\rmatI}{\bimI}
\safemath{\rmatJ}{\bimJ}
\safemath{\rmatK}{\bimK}
\safemath{\rmatL}{\bimL}
\safemath{\rmatM}{\bimM}
\safemath{\rmatN}{\bimN}
\safemath{\rmatO}{\bimO}
\safemath{\rmatP}{\bimP}
\safemath{\rmatQ}{\bimQ}
\safemath{\rmatR}{\bimR}
\safemath{\rmatS}{\bimS}
\safemath{\rmatT}{\bimT}
\safemath{\rmatU}{\bimU}
\safemath{\rmatV}{\bimV}
\safemath{\rmatW}{\bimW}
\safemath{\rmatX}{\bimX}
\safemath{\rmatY}{\bimY}
\safemath{\rmatZ}{\bimZ}

\safemath{\rmatDelta}{\bimDelta}
\safemath{\rmatLambda}{\bimLambda}
\safemath{\rmatPhi}{\bimPhi}
\safemath{\rmatSigma}{\bimSigma}
\safemath{\rmatOmega}{\bimOmega}
\safemath{\rmatTheta}{\bimTheta}   

\usepackage{amssymb}
\usepackage{amsfonts}
\usepackage{mathrsfs}
\usepackage{xspace}
\usepackage{bm}
\usepackage{fancyref}
\usepackage{textcomp} 
\usepackage{color}

\newenvironment{textbmatrix}{	\setlength{\arraycolsep}{2.5pt}%
								\big[\begin{matrix}}{\end{matrix}\big]%
								\raisebox{0.08ex}{\vphantom{M}}}   
								
\def\be{\begin{equation}}
\def\ee{\end{equation}}
\def\een{\nonumber \end{equation}}
\def\mat{\begin{bmatrix}}
\def\emat{\end{bmatrix}}
\def\btm{\begin{textbmatrix}}
\def\etm{\end{textbmatrix}}

\def\ba#1\ea{\begin{align}#1\end{align}}
\def\bas#1\eas{\begin{align*}#1\end{align*}}
\def\bs#1\es{\begin{split}#1\end{split}} 
\def\bg#1\eg{\begin{gather}#1\end{gather}}
\def\bml#1\eml{\begin{multline}#1\end{multline}}
\def\bi#1\ei{\begin{itemize}#1\end{itemize}}

\newcommand{\subtext}[1]{\text{\fontfamily{cmr}\fontshape{n}\fontseries{m}\selectfont{}#1}}

\newcommand{\sub}[1]{\ensuremath{_{\subtext{#1}}}} 

\DeclareMathOperator{\had}{\odot}			
\DeclareMathOperator{\four}{\opF}			

\DeclareMathOperator{\landauO}{\mathcal{O}}

\newcommand{\Ex}[2]{\ensuremath{\Exop_{#1}\mathopen{}\left[#2\right]}} 	
\newcommand{\abs}[1]{\left\lvert#1\right\rvert}		

\newcommand{\vecnorm}[1]{\lVert#1\rVert}		
\newcommand{\conj}[1]{\ensuremath{#1^{*}}} 	
\newcommand{\tp}[1]{\ensuremath{#1^{T}}} 		
\newcommand{\herm}[1]{\ensuremath{#1^{H}}} 	

\safemath{\dirac}{\delta}					
\safemath{\krond}{\dirac}					
\newcommand{\allz}[2]{\ensuremath{#1=0,1,\ldots,#2-1}}

\newcommand{\logdet}[1]{\log \det\mathopen{}\left(#1\right)} 
\safemath{\upto}{\uparrow}
\safemath{\downto}{\downarrow}
\safemath{\iu}{j}							
\safemath{\ev}{\lambda}						
\safemath{\hilseqspace}{l^{2}}				
\newcommand{\banachfunspace}[1]{\setL^{#1}}	
\safemath{\hilfunspace}{\banachfunspace{2}}	
       
\DeclareMathOperator{\diag}{diag}      		   

\safemath{\snr}{\rho} 				
\safemath{\No}{N_0}							
\safemath{\Es}{E_s}							
\safemath{\Eb}{E_b}							
\safemath{\EbNo}{\frac{\Eb}{\No}}
\safemath{\EsNo}{\frac{\Es}{\No}}

\DeclareMathOperator{\CHop}{\ensuremath{\opH}} 
\safemath{\tvir}{\rndh_{\CHop}}				
\safemath{\tvtf}{\rndl_{\CHop}}				
\safemath{\spf}{\rnds_{\CHop}}				
\safemath{\bff}{H_{\CHop}}					

\safemath{\ircf}{r_{h}}						
\safemath{\tftvcf}{r_{s}}					
\safemath{\tfcf}{r_{l}}						
\safemath{\bfcf}{r_{H}}						

\safemath{\tcorr}{c_h}						
\safemath{\scf}{c_{s}}						
\safemath{\tfcorr}{c_{l}}					
\safemath{\fcorr}{c_{H}}						

\safemath{\mi}{I}							
\safemath{\capacity}{C}						
\safemath{\difent}{h}						

\newcommand{\iid}{i.i.d.\@\xspace}

\safemath{\normal}{\mathcal{N}}			
\safemath{\jpg}{\mathcal{CN}}			
\safemath{\mchain}{\leftrightarrow}		
\newcommand{\given}{\,\vert\,}				
  
\safemath{\wpone}{\text{w.p.1}}		

\safemath{\dB}{\,\mathrm{dB}}
\safemath{\dBm}{\,\mathrm{dBm}}
\safemath{\Hz}{\,\mathrm{Hz}}
\safemath{\kHz}{\,\mathrm{kHz}}
\safemath{\MHz}{\,\mathrm{MHz}}
\safemath{\GHz}{\,\mathrm{GHz}}
\safemath{\s}{\,\mathrm{s}}
\safemath{\ms}{\,\mathrm{ms}}
\safemath{\mus}{\,\mathrm{\text{\textmu}s}}
\safemath{\ns}{\,\mathrm{ns}}
\safemath{\ps}{\,\mathrm{ps}}
\safemath{\meter}{\,\mathrm{m}}
\safemath{\mm}{\,\mathrm{mm}}
\safemath{\cm}{\,\mathrm{cm}}
\safemath{\m}{\,\mathrm{m}}
\safemath{\W}{\,\mathrm{W}}
\safemath{\mW}{\, \mathrm{mW}}
\safemath{\J}{\,\mathrm{J}}
\safemath{\K}{\,\mathrm{K}}
\safemath{\bit}{\,\mathrm{bit}}
\safemath{\nat}{\,\mathrm{nat}} 
\safemath{\kg}{\,\mathrm{kg}} 

\safemath{\define}{\triangleq}			

\newcommand{\sothat}{\,:\,}				
\providecommand{\inner}[2]{\ensuremath{\langle#1,#2\rangle}}
\safemath{\equivalent}{\sim}
\safemath{\distas}{\sim}					
\safemath{\sdiff}{\Delta}				

\safemath{\reals}{\mathbb{R}}
\safemath{\positivereals}{\reals_{+}}
\safemath{\integers}{\mathbb{Z}}
\safemath{\posint}{\integers_{+}}
\safemath{\naturals}{\mathbb{N}}
\safemath{\posnaturals}{\naturals_{+}}
\safemath{\complexset}{\mathbb{C}}
\safemath{\rationals}{\mathbb{Q}}

\newcommand*{\fancyrefapplabelprefix}{app}		
\newcommand*{\fancyrefthmlabelprefix}{thm}		
\newcommand*{\fancyreflemlabelprefix}{lem}		
\newcommand*{\fancyrefcorlabelprefix}{cor}		
\newcommand*{\fancyrefdeflabelprefix}{def}		
\newcommand*{\fancyrefproplabelprefix}{prop}		
\newcommand*{\fancyrefexelabelprefix}{exe}		
\frefformat{vario}{\fancyrefseclabelprefix}{Section~#1}
\frefformat{vario}{\fancyrefthmlabelprefix}{Theorem~#1}
\frefformat{vario}{\fancyreflemlabelprefix}{Lemma~#1}
\frefformat{vario}{\fancyrefcorlabelprefix}{Corollary~#1}
\frefformat{vario}{\fancyrefdeflabelprefix}{Definition~#1}
\frefformat{vario}{\fancyreffiglabelprefix}{Fig.~#1}
\frefformat{vario}{\fancyrefapplabelprefix}{Appendix~#1}
\frefformat{vario}{\fancyrefeqlabelprefix}{(#1)}
\frefformat{vario}{\fancyrefproplabelprefix}{Property~#1}
\frefformat{vario}{\fancyrefexelabelprefix}{Exercise~#1}

\let\difent\undefined
\safemath{\difent}{\mathsf{h}}
\newcommand{\difentp}[1]{\difent\mathopen{}\left(#1\right)}
 \safemath{\uncertainty}{\epsilon} 

\newcommand{\specdensity}[1]{c_{#1}}			
\newcommand{\specdensitymat}[1]{\matC_{#1}}	

\let\time\undefined
\safemath{\bandwidth}{W}		
\safemath{\duration}{D}			
\safemath{\time}{t}				
\safemath{\freq}{f}				
\safemath{\dtdf}{\dtime,\dfreq}  
\safemath{\altdtdf}{\altdtime,\altdfreq}
\safemath{\dtime}{k}				
\safemath{\dfreq}{n}				
\safemath{\altdtime}{l}				
\safemath{\altdfreq}{m}				

\safemath{\dtimetilde}{\tilde{\dtime}}				
\safemath{\dfreqtilde}{\tilde{\dfreq}}				
\safemath{\altdtimetilde}{\tilde{\altdtime}}				
\safemath{\altdfreqtilde}{\tilde{\altdfreq}}				

\safemath{\Ddtime}{\dtime}		
\safemath{\Ddfreq}{\dfreq}		
\safemath{\delay}{\tau}			
\safemath{\doppler}{\nu}			
\safemath{\maxDoppler}{\doppler_{0}}		
\safemath{\altmaxDoppler}{\widetilde{\doppler}_0}
\safemath{\maxDelay}{\delay_{0}}			
\safemath{\altmaxDelay}{\widetilde{\delay}_{0}}
\safemath{\spread}{\Delta_{\CHop}}		
\safemath{\altspread}{\widetilde{\Delta}_{\CHop}}
\safemath{\tstep}{T}				
\safemath{\fstep}{F}				
\safemath{\alttstep}{\widetilde{\tstep}}				
\safemath{\altfstep}{\widetilde{\fstep}}				
\safemath{\tfstep}{\tstep\fstep}	
\safemath{\srtfstep}{\sqrt{\tfstep}}
\safemath{\fslots}{N}			
\safemath{\tslots}{K}			
\safemath{\tslotstot}{\tslots_x}
\safemath{\fslotstot}{\fslots_x}
\safemath{\tslotsguard}{K_g}
\safemath{\tfslots}{\tslots\fslots}	
\safemath{\tfsamples}{\dtime\tstep,\dfreq\fstep}	
\safemath{\WHset}{(\logon,\tstep,\fstep)}
\safemath{\WHsetsquare}{(\logon,\srtfstep,\srtfstep)}
\safemath{\spreadset}{\setD}
\safemath{\altspreadset}{\setE}
\safemath{\spreadsquareset}{\widetilde{\spreadset}}

\safemath{\dd}{\delay,\doppler}
\let\tvir\undefined
\safemath{\tvir}{h_{\CHop}}				
\safemath{\tvirp}{\tvir(\time,\delay)}	
\let\spf\undefined
\safemath{\spf}{S_{\CHop}}				
\safemath{\spfp}{\spf(\delay,\doppler)}	
\safemath{\corrtd}{R_{\CHop}}			
\safemath{\corrtf}{B_{\CHop}}			
\safemath{\mvchcovmat}{\matR{}}	
\safemath{\corrtdp}{\corrtd(\time,\delay)}
\safemath{\scafun}{C_{\CHop}}			
\safemath{\scafunp}{\scafun(\delay,\doppler)}	
\safemath{\ch}{h}						
\safemath{\altch}{\tilde{h}}					
\safemath{\wgn}{w}						
\safemath{\wgnp}{\wgn(\time)}			
\safemath{\wgnvec}{\vecw}				
\safemath{\chvec}{\vech}					
\safemath{\specparam}{\theta}			
\safemath{\altspecparam}{\varphi}		
\safemath{\chcorr}{r}					
\safemath{\chcorrp}{\chcorr(\time,\freq)}    
\safemath{\chspecfun}{\specdensity{}}		
\safemath{\chspecfunp}{\specdensity{}(\altspecparam,\specparam)} 
\safemath{\chspecfunmat}{\specdensitymat{}}	
\safemath{\chspecfunmatp}{\chspecfunmat(\specparam)}	
\safemath{\chspecfunentry}{c}			
\safemath{\chspecfunentryp}{\chspecfunentry_{\dfreq}(\specparam)}
\safemath{\chspecfunentrypzero}{\chspecfunentry_{0}(\specparam)}
\safemath{\approxeig}{c}				
\safemath{\approxeigp}{\approxeig[\dtdf]}

\safemath{\chset}{\setH}
\safemath{\chsetp}{\chset\mathopen{}\left(\maxDelay,\maxDoppler,\uncertainty\right)} 
\safemath{\innrate}{\lambda}

\safemath{\logon}{g}						
\safemath{\logonp}{\logon(\time)}
\safemath{\altlogon}{\widetilde{\logon}}						
\safemath{\altaltlogon}{e}
\safemath{\altlogonp}{\altlogon(\time)}
\safemath{\altaltlogonp}{\altaltlogon(\time)}
\safemath{\slogon}{\logon_{\dtime,\dfreq}}	
\safemath{\slogonp}{\slogon(\time)}	
\safemath{\slogonset}{\left\{\logon(\time-\dtime\srtfstep)\cex{\dfreq\srtfstep\time}\right\}}
\safemath{\eftime}{T_{0}}	
 \safemath{\efband}{F_{0}} 
 \safemath{\logonf}{G} 
 \safemath{\logonfp}{\logonf(\freq)}
 \safemath{\logonalt}{f}
 \safemath{\logonaltp}{\logonalt(\time)}
 \safemath{\slogonct}{\logon_{(\alpha,\beta)}}
 \safemath{\slogonctalt}{\logon_{(\alpha',\beta')}}
 \safemath{\af}{A}						
 \safemath{\afp}{\af_{\logon}(\delay,\doppler)}	
\safemath{\setloc}{\setG}
\safemath{\setloclogon}{\widetilde{\setG}}
\safemath{\dertau}{b}
\safemath{\dernu}{a}
\safemath{\dertaup}{\dertau_{\dtdf}}
\safemath{\dernup}{\dernu_{\dtdf}}

\safemath{\const}{c}
\safemath{\constM}{\const_{M}}
\safemath{\constm}{\const_{m}}

\safemath{\corr}{d}
\safemath{\corrmat}{\matD}

\safemath{\timedecay}{\time_0}
\safemath{\decay}{\mu}
\safemath{\decayalt}{\decay'}
\safemath{\constdecay}{\gamma}
\safemath{\constdecayalt}{\constdecay'}
\safemath{\dur}{\zeta}
\safemath{\rolloff}{\delta}
\safemath{\auxfun}{S}

\safemath{\inp}{x}				
\safemath{\inpp}{\inp(\time)}	
\safemath{\inpfour}{X}			
\safemath{\inpfourp}{\inpfour(\freq)}	
\safemath{\outp}{y}				
\safemath{\outpfilt}{\outp_f}
\safemath{\outpfiltp}{\outpfilt(\time)}
\safemath{\outpp}{\outp(\time)} 
\safemath{\inpvec}{\vecx}			
\safemath{\inpvecpred}{\inpvec_{\text{prec}}^{(\dtdf)}}
\safemath{\outpvec}{\vecy}		
\safemath{\outpnn}{r}			
\safemath{\outpnnp}{\outpnn(\time)}	
\safemath{\outpnnfilt}{\outpnn_f}
\safemath{\outpnnfiltp}{\outpnnfilt(\time)}
\safemath{\Pave}{P}				
\safemath{\avP}{\Ex{}{\vecnorm{\inpvec}^{2}}\le\tslots\Pave\,\tstep}
\safemath{\avPnorm}{(1/\tstep)\Ex{}{\vecnorm{\inpvec}^{2}}\le\tslots\Pave}
\safemath{\setinp}{\setQ}		
\safemath{\probinpdisc}{\setinp_d}
\safemath{\SNR}{\rho}			

\safemath{\miconttime}{\mi_{\duration}}
\safemath{\capacitydisc}{\capacity_d}
\safemath{\probinp}{\setQ_\inp}
\safemath{\proboutpf}{\setQ_{\outpnnfilt}}
\safemath{\probinpset}{\setQ(\bandwidth,\duration,\spillover,\Pave)}       
\safemath{\altprobinpset}{\widetilde{\setQ}} 
\safemath{\probinpwhset}{\setQ\sub{WH}(\bandwidth,\duration,\spillover,\Pave)}  
 \safemath{\capacityp}{\capacity(\SNR)}	
\safemath{\minambsquare}{m_{\logon}}
\safemath{\sumambsquare}{M}
\safemath{\sumambsquarep}{\sumambsquare(\dd)}
\safemath{\maxsumambsquare}{M_\logon}
\safemath{\LB}{L_1}
\safemath{\LBsimple}{L_2}
\safemath{\LBsimplep}{\LBsimple(\SNR,\logonp,\tstep,\fstep,\maxDelay,\maxDoppler,\uncertainty)}
\safemath{\LBsquare}{\LBsimple^{(s)}}
\safemath{\LBsquarep}{\LBsquare\mathopen{}\left(\SNR,\altlogonp,\tfstep,\spread,\uncertainty\right)}
\safemath{\LBsquarepnew}{\LBsquare\mathopen{}\left(\SNR,\logonp,\tfstep,\spread,\uncertainty\right)}
\safemath{\LBund}{\L_{u}}
\safemath{\LBunds}{\widetilde{L}_{u}}
\safemath{\LBundp}{\LBund(\SNR,\logon,\tfstep,\spread,\uncertainty)}
\safemath{\LBundsp}{\LBunds(\SNR,\logon,\tfstep,\spread,\uncertainty)}
\safemath{\ccoh}{\capacity_{coh}} 
\safemath{\cawgn}{\capacity_{\text{AWGN}}} 
\safemath{\cawgnp}{\cawgn(\SNR)} 

\safemath{\ccohp}{\ccoh(\SNR)} 
\safemath{\SNRmax}{\SNR_{\text{max}}}
\safemath{\SNRmin}{\SNR_{\text{min}}}
\safemath{\fun}{f}
\safemath{\funp}{\fun(\SNR)}
\safemath{\SNRnorm}{\gamma}
\safemath{\SNRnormMax}{\SNRnorm_{max}}
\safemath{\SNRnormMin}{\SNRnorm_{min}}

\safemath{\sumdtime}{\sum_{\dtime=-\infty}^{\infty}} 
\safemath{\sumdfreq}{\sum_{\dfreq=-\infty}^{\infty}} 

\safemath{\sumdtimelimited}{\sum_{\dtime=-\tslots}^{\tslots}} 
\safemath{\sumdfreqlimited}{\sum_{\dfreq=-\fslots}^{\fslots}} 

\safemath{\altsumdtimelimited}{\sum_{\altdtime=-\tslots}^{\tslots}} 
\safemath{\altsumdfreqlimited}{\sum_{\altdfreq=-\fslots}^{\fslots}} 

 \renewcommand{\setminus}{\backslash}						
 \safemath{\limintime}{\lim_{\tslotstot\to\infty}}	
 \safemath{\imat}{\matidentity}
\newcommand{\spreadint}[1]{\iint_{\doppler\,\delay} #1 d\delay d\doppler}

\newcommand{\intspecparamin}[1]{\int_{\specparam\in \setB} #1 d\specparam}
\newcommand{\intspecparamout}[1]{\int_{\specparam \in \bar{\setB}} #1 d\specparam}					
 \newcommand{\cex}[1]{e^{\iu2\pi #1}}				
 \newcommand{\cexn}[1]{e^{-\iu2\pi #1}}			
\safemath{\euler}{\gamma}	
\safemath{\intdiscrete}{\int_{-1/2}^{1/2}}
\safemath{\intdiscretetwod}{\intdiscrete\intdiscrete}
\safemath{\hilfunspacep}{\hilfunspace(\reals)}	
\safemath{\hilfuntimeband}{\hilfunspace(\bandwidth,\duration,\spillover)}
\safemath{\spillover}{\eta}
\safemath{\hilfunband}{\hilfunspace_\bandwidth(\reals)}
\newcommand{\timetrunc}[1]{\mathbb{T}_{#1}}	
\newcommand{\freqtrunc}[1]{\mathbb{B}_{#1}}	
\safemath{\onb}{\phi}
\safemath{\indexonb}{m}
\safemath{\numonb}{M}
\safemath{\onbp}{\onb_{\indexonb}(\time)}
\safemath{\funloc}{f}
\safemath{\funlocp}{\funloc(\dd)}
\newcommand{\eigmax}[1]{\lambda_{\text{max}}\{#1\}}
\safemath{\ofun}{\phi}
\safemath{\altofun}{\ofun'}
\safemath{\oindex}{m}
\safemath{\ofunp}{\ofun_\oindex(\time)}
\safemath{\altofunp}{\altofun_\oindex(\time)}
\safemath{\onum}{M}        

\newcommand{\filtinp}{\inp_f^{(\delay,\doppler)}}
\newtheorem{thm}{Theorem}
\newtheorem{lem}[thm]{Lemma}

\newtheorem{cor}[thm]{Corollary}
\newtheorem{dfn}[thm]{Definition}
\newtheorem{prop}{Property}

\safemath{\uncerMax}{\uncertainty^{max}}
\safemath{\spreadMax}{\spread^{max}}

 \safemath{\interf}{p}						
 \safemath{\interfp}{\interf[\altdtdf,\dtdf]}
 \safemath{\interfmat}{\matP}						
 \safemath{\intvar}{\sigma^2_{\interf}}
 \safemath{\intvarp}{\intvar[\altdtime-\dtime,\altdfreq-\dfreq]}
 \safemath{\altwgn}{\widetilde{\wgn}}
 \safemath{\altwgnp}{\altwgn[\dtdf]}
 \safemath{\altwgnpgauss}{\altwgn_{G}[\dtdf]}
\safemath{\intvartot}{\sigma^{2}_{I}}
 \safemath{\wgnvecone}{\wgnvec_{1}}				
 \safemath{\wgnvectwo}{\wgnvec_{2}}				
 \safemath{\outpvecone}{\outpvec_{1}}                               
 \safemath{\outpvectwo}{\outpvec_{2}}                               
 \safemath{\noisesplit}{\alpha}                                              
 \safemath{\corrintinp}{\matK(\inpvec)}

\safemath{\measure}{\mu}
\safemath{\psdoned}{c}
\safemath{\psdonedp}{\psdoned(\specparam)}
\safemath{\spreadoned}{\Delta}

\begin{document}

\title{On the Sensitivity of Continuous-Time Noncoherent Fading Channel Capacity}

\author{Giuseppe Durisi,~\IEEEmembership{Senior Member,~IEEE}, Veniamin I. Morgenshtern, and\\ Helmut B\"olcskei,~\IEEEmembership{Fellow,~IEEE}
\thanks{G. Durisi is with the Department of Signals and Systems, Chalmers University of Technology, Gothenburg, Sweden, Email: durisi@chalmers.se} 
\thanks{V. I. Morgenshtern is with the Department of Statistics, Stanford University, CA, USA, Email: vmorgen@stanford.edu}
\thanks{H. B\"olcskei is with the Department of Information Technology and Electrical Engineering, ETH Zurich, Zurich, Switzerland, Email: boelcskei@nari.ee.ethz.ch}
\thanks{Part of the material in this paper was presented at the 2009 IEEE International Symposium on Information Theory.} 
\thanks{ Copyright (c) 2012 IEEE. Personal use of this material is permitted. ÊHowever, permission to use this material for any other purposes must be obtained from the IEEE by sending a request to pubs-permissions@ieee.org}
}

\maketitle

\begin{abstract}
The noncoherent capacity of stationary discrete-time fading channels is known to be very sensitive to the fine details of the channel model. 
More specifically, the measure of the support of the fading-process power spectral density (PSD) determines if noncoherent capacity grows logarithmically in SNR or slower than logarithmically. 
Such a result is unsatisfactory from an engineering point of view, as the support of the PSD cannot be determined through measurements.
The aim of this paper is to assess whether, for general continuous-time Rayleigh-fading channels, this sensitivity has a noticeable impact on capacity at SNR values of practical interest. 

To this end, we consider the general class of band-limited continuous-time Rayleigh-fading channels that satisfy the wide-sense stationary uncorrelated-scattering (WSSUS) assumption and are, in addition, underspread. 
We show that,  for all SNR values of practical interest, the noncoherent capacity of 
every channel in this class is close to the capacity of an AWGN channel with the same SNR and bandwidth, independently of the measure of the support of the scattering function (the two-dimensional channel PSD).  
Our result is based on a lower bound on noncoherent capacity, which is built on a  discretization of the channel input-output relation induced by projecting  onto Weyl-Heisenberg (WH) sets.
This approach is interesting in its own right as it yields a mathematically tractable way of dealing with the mutual information between certain continuous-time random signals.
\end{abstract}

\section{Introduction and Summary of Results}
\label{sec:introduction}
The capacity of fading channels in the \emph{noncoherent setting} where neither transmitter nor  receiver are aware of the realizations of the fading process, but both know its statistics,\footnote{Capacity in the noncoherent setting is sometimes called \emph{noncoherent} capacity; in the remainder of this paper, it  will be referred to simply as capacity.
We will use the adjective \emph{coherent} to denote the setting where the channel realizations are perfectly known at the receiver but unknown at the transmitter, which is assumed to know the channel statistics only.} is  notoriously difficult to analyze, even for simple channel models.
Most of the results available in the literature pertain to either low or high signal-to-noise ratio (SNR) asymptotics. 
While in the low-SNR regime the capacity behavior is robust with respect to the underlying channel model (see for example~\cite{durisi10-01a,durisi11-03b}), this is not the case in the high-SNR regime, where---as we are going to argue next---capacity is very sensitive to the \emph{fine details} of the channel model.

Consider, e.g., a \emph{discrete-time} stationary frequency-flat time-selective Rayleigh-fading channel subject to additive white Gaussian noise (AWGN). 
Here, the channel statistics are fully specified by  the fading-process power spectral density (PSD)~$\psdonedp$, $\specparam \in [-1/2,1/2)$, and by the noise variance. 
The high-SNR capacity of this channel turns out to depend on the measure~\measure of the support of the PSD. 
More specifically, let~\SNR denote the SNR; if $\measure<1$, capacity behaves as~$(1-\measure)\log \SNR$ in the high-SNR regime~\cite{lapidoth05-07a}. 
The \emph{pre-log} factor $(1-\measure)$ quantifies the loss in signal-space dimensions (relative to coherent capacity~\cite{biglieri98-10a}, which behaves  as $\log \SNR$) due to the lack of channel knowledge at the receiver.\footnote{{Results of the same nature as those reported in~\cite{lapidoth05-07a}  were  obtained previously for the block-fading channel model (a non-stationary channel model) in~\cite{zheng02-02a,hassibi03-04a}}.} 
For $\measure\ll 1$ this loss is negligible, suggesting that, in this case, the  realizations of the fading channel can be learned at the receiver (at high SNR) by sacrificing a negligible fraction of the signal-space dimensions available for communication.
If~$\measure=1$ and the fading process is \emph{regular}, i.e.,~$\int_{-1/2}^{1/2} \log \psdonedp d\specparam> -\infty$,  the high-SNR capacity behaves as~$\log\log \SNR$~\cite{lapidoth03-10a}. 
This double-logarithmic growth behavior of capacity with SNR renders communication in the high-SNR regime extremely power inefficient.

\begin{figure}[t]
	\centering
		\includegraphics[width=\figwidth]{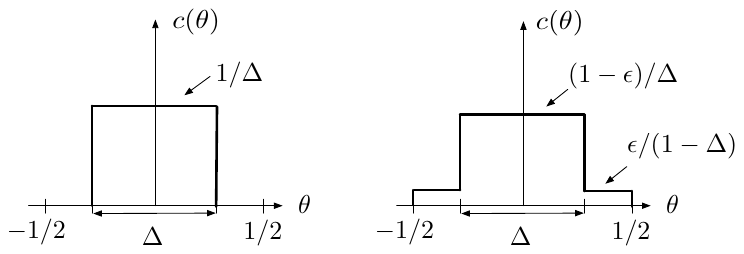}
	\caption{Two channels with similar PSD $c(\theta)$, but drastically different high-SNR capacity behavior.}
	\label{fig:psdrect}
\end{figure}

As a consequence of the results just mentioned, we have the following: consider two discrete-time stationary Rayleigh-fading channels, the first one with PSD equal to~$1/\spreadoned$ for~$\specparam \in [-\spreadoned/2,\spreadoned/2]$  and~$0$ else ($0<\spreadoned<1$), and the second one with PSD equal to~$(1-\uncertainty)/\spreadoned$ for~$\specparam \in [-\spreadoned/2,\spreadoned/2]$ and~$\uncertainty/(1-\spreadoned)$ else ($0<\uncertainty<1$, see  \fref{fig:psdrect}).
These two channels will have completely different high-SNR capacity behavior, no matter how small~$\uncertainty$ is.   
Specifically, the capacity of the first channel behaves as $(1-\spreadoned)\log \SNR$, whereas the capacity of the second one grows as $\log\log\SNR$.
A result like this is clearly unsatisfactory from an engineering point of view, as the measure of the support of a PSD cannot be determined through channel measurements.
Such a sensitive dependency of the (high-SNR) capacity behavior on the fine details of the channel model (by fine details we mean details that, in the words of Slepian~\cite{slepian76-03a}, have ``\dots no direct meaningful counterparts in the real world \dots ''), should make one question the usefulness of the discrete-time stationary channel model itself, at least for high-SNR analyses.
In the light of this observation, an engineering-relevant problem is  to 
assess whether this sensitivity has a noticeable impact on  capacity at SNR values of practical interest.
Unfortunately, this problem is still largely open.
For the stationary discrete-time case, an attempt to characterize the capacity sensitivity was made in~\cite{etkin06-04a},  where, for a first-order Gauss-Markov channel process
 (a regular process), the SNR beyond which capacity starts exhibiting a sub-logarithmic growth in SNR is computed as a function of the innovation variance $\innrate$ of the process.
More specifically, it is shown in~\cite{etkin06-04a} that for $\snr \gg 1$ and $\lambda \ll 1$ capacity grows as $\log \snr$ as long as $\snr < 1/\lambda$.
In words, when the innovation variance is small, the high-SNR capacity grows logarithmically in SNR up to SNR values not exceeding $1/\lambda$. 
The main limitation of this result lies in the fact that it is based on a highly specific channel model, namely a first-order Gauss-Markov process, which is fully described by a single parameter, the innovation variance.
Furthermore, it is difficult to relate this parameter to physical channel quantities such as the channel Doppler spread.

A more general approach is presented in~\cite{lapidoth03-10a}, where the \emph{fading number}, defined as the second term in the high-SNR expansion of capacity, is characterized for arbitrary discrete-time, stationary, regular fading channels.
The fading number determines the rate after which the $\log\log$ regime kicks in, and communication becomes extremely power inefficient.
Unfortunately, as illustrated in~\cite{koch05-02a}, it is, in general, not possible to relate the fading number to the SNR value at which the $\log\log$ behavior comes into effect.

The purpose of this paper is to characterize the sensitivity of capacity with respect to the channel model for the
general class of \emph{continuous-time} Rayleigh-fading linear time-varying (LTV) channels that satisfy the \emph{wide-sense stationary} (WSS) and \emph{uncorrelated scattering} (US) assumptions~\cite{bello63-12a} and that are, in addition, \emph{underspread}~\cite{kennedy69}. 
The Rayleigh-fading and the WSSUS assumptions imply that the statistics of the channel are fully characterized by its two-dimensional PSD, often referred to as the \emph{scattering function}~\cite{bello63-12a}; the underspread assumption is satisfied if the scattering function is ``highly concentrated'' in the delay-Doppler plane.  
Different definitions of the underspread property are available in the literature (e.g., in terms of the support area of the scattering function~\cite{durisi10-01a,kozek97-03a} or in terms of its moments~\cite{matz03a}).      
For the problem considered in this paper, it is crucial to adopt a novel definition  of the underspread property (see Definition~\ref{def:underspread} in \fref{sec:a_robust_definition_of_underspread_channels}), inspired by Slepian's treatment of finite-energy signals that are approximately time- and band-limited~\cite{slepian76-03a}.
Specifically, we shall say that a WSSUS channel is underspread if  its scattering function has only a fraction~$\uncertainty\ll 1$ of its volume outside a rectangle of area~$\spread \ll 1$ . 
This novel definition of the underspread property encompasses the  underspread definitions previously proposed in the literature~\cite{kozek97-03a,durisi10-01a,matz03a} and generalizes them. 

When $\uncertainty=0$, i.e., when the scattering function is compactly supported, and $\spread\ll 1$ we expect---on the basis of the results obtained in~\cite{lapidoth03-10a,lapidoth05-07a} in the context of the stationary discrete-time fading channel model---capacity to grow logarithmically in SNR.
Unfortunately, it is not possible to determine through channel measurements whether a scattering function is compactly supported or not, which motivates our novel underspread definition.
For the practically more relevant case $0<\uncertainty\ll 1$, we show that the sub-logarithmic growth behavior kicks in only at very large SNR. 
Our result is built on a  lower bound on the capacity of band-limited continuous-time WSSUS underspread Rayleigh-fading channels that is explicit in the channel parameters~\spread and~\uncertainty.
By comparing this lower bound to a trivial capacity upper bound, namely, the capacity of a nonfading AWGN channel with the same
SNR and bandwidth, we find that, for all SNR values of practical interest, 
the fading channel capacity is close\footnote{{``Close'' here means that the ratio between the capacity lower bound  and the capacity of a nonfading AWGN channel (with the same SNR and bandwidth) exceeds $0.75$.}} to the capacity of a nonfading AWGN channel (with the same SNR and bandwidth).                                
As a rule of thumb, this statement is true for all SNR values in the range  $\sqrt{\spread}\ll \SNR \ll 1/(\spread+\epsilon)$.
Hence, we conclude that  the fading channel capacity essentially grows logarithmically in SNR for all SNR values of practical interest.

Information theoretic analyses of continuous-time channels are notoriously difficult.
The standard approach is to discretize the continuous-time channel input-output (I/O) relation by projecting the input and output signals onto the singular functions of the \emph{channel operator}~\cite{wyner66-03a,gallager68a}. 
This  yields a \emph{diagonalized} discretized I/O relation consisting of countably many scalar, non-interacting I/O relations. 
Unfortunately, this approach is not viable in our setting because random LTV channels have random singular functions, which are not known to transmitter and receiver in the noncoherent setting~\cite{durisi10-01a,durisi11-03b}. 
We will nevertheless discretize the channel by constraining the input signal to lie in the span of an orthonormal Weyl-Heisenberg (WH) set, i.e., a set of time-frequency shifted versions of a given function, and by projecting the receive signal on the same set of functions.
This guarantees that the resulting discretized channel inherits the (two-dimensional) stationarity property of the underlying continuous-time channel, a fact that is essential for our analysis.  
This approach is interesting in its own right, as it yields a mathematically tractable way of dealing with the mutual information between certain continuous-time random signals.

In~\cite{durisi10-01a} a similar approach was used to obtain bounds on the capacity of continuous-time Rayleigh-fading WSSUS underspread channels at low SNR.
These bounds are derived under the assumption that the  off-diagonal terms in the discretized I/O relation can be neglected, which greatly simplifies the capacity analysis.  
Whereas this simplification was shown in~\cite{durisi11-03b} to be admissible at low SNR, it is unclear whether the off-diagonal terms can be neglected at high SNR.                                                                               
We will therefore explicitly account for the off-diagonal terms in the discretized I/O relation by treating them as (signal-dependent) additive noise, and thus obtain a firm lower bound on the capacity of the underlying continuous-time channel.
This lower bound yields an information-theoretic criterion for the design of WH sets to be used for pulse-shaped (PS) orthogonal frequency-division multiplexing (OFDM) communication systems operating over  Rayleigh-fading WSSUS underspread fading channels.                                                                                                               
In particular, the lower bound suggests that the WH set should be chosen so as to optimally trade signal-space dimensions (available for communication) for minimization of the power of the off-diagonal terms in the resulting discretized I/O relation.
\subsubsection*{Notation}
 Uppercase boldface letters denote matrices, and lowercase boldface letters
 designate vectors. 
The Hilbert space of complex-valued finite-energy signals is denoted as~\hilfunspacep; furthermore,
$\inner{\cdot}{\cdot}$ and~$\vecnorm{\cdot}$ stand for the inner product and the norm in~\hilfunspacep, respectively.
 The set of positive real numbers is denoted as~\positivereals and the set of integers as \integers; 
$\Ex{}{\cdot}$ is the expectation operator, $\difent(\cdot)$ denotes differential entropy, and~$\four[\cdot]$ stands for the Fourier transform. 
For two vectors~\veca and~\vecb of equal dimension, the Hadamard (element-wise) product is denoted as~$\veca \had \vecb$.
We write~$\diag\{\inpvec\}$ for the diagonal matrix that has the elements of the vector~\inpvec on its main diagonal.   
The superscripts~$\tp{}$, $\conj{}$, and~$\herm{}$ stand for transposition,
element-wise conjugation, and Hermitian transposition, respectively.
The largest eigenvalue of a Hermitian matrix~\matA is denoted as~$\eigmax{\matA}$.  
For two functions~$f(x)$ and~$g(x)$, the notation~$f(x)=\landauO(g(x))$, $x\to \infty$, means that $\lim\sup_{x\to \infty}\abs{f(x)/g(x)}<\infty$. 
Finally,~$\krond[k]$ is defined as~$\krond[0]=1$ and~$\krond[k]=0$ for~$k\neq 0$.
Throughout the paper, we shall make use of the following projection operators acting on $\hilfunspace(\reals)$:
the \emph{time-limiting} operator $\timetrunc{\duration}$, defined as
\begin{equation*}
	(\timetrunc{\duration}\inp)(\time) =	\begin{cases}
							\inpp, & \text{if } \abs{\time}\leq \duration/2\\
							0,	  & \text{otherwise}
						\end{cases}
\end{equation*}
which limits~\inpp  to the interval $[-\duration/2,\duration/2]$,
and the \emph{frequency-limiting} operator defined as
\begin{equation*}
	(\freqtrunc{\bandwidth}\inp)(\time)=\int_{\time'} \frac{\sin[\pi\bandwidth(\time-\time')]}{\pi(\time-\time')}\inp(\time')d\time'
\end{equation*}
which limits the Fourier transform of  \inpp  to the interval $[-\bandwidth/2,\bandwidth/2]$.

\section{System Model} 
\label{sec:system_model}
\subsection{Channel and Signal Model} 
\label{sec:the_continuous_time_input_output_relation}
%
%
%
The  I/O relation of a continuous-time random LTV channel~$\CHop$ can be written as~\cite{matz11-03a}
\begin{align}
	\outp(\time) &= \underbrace{(\CHop\inp)(\time)}_{\define\, \outpnnp} +\,\wgn(\time) \notag \\
	&=\int_{\delay}\!\tvirp\inp(\time-\delay)d\delay +\wgn(\time).
\label{eq:ltv-io}
\end{align}
Here,~\outpnnp is the output signal in the absence of additive noise.
Following~\cite[Model 2]{wyner66-03a}, we assume that  the stochastic input signal~\inpp:
\begin{enumerate}[i)]
	\item 
	is strictly band-limited to $\bandwidth \Hz$ according to
	\begin{equation}\label{eq:bandwidth_constraint}
	\inpfourp=0, \quad \text{for} \abs{\freq}> \bandwidth/2	
	\end{equation}
	with probability one, where $\inpfourp\define\four[\inpp]$;
	\item  is approximately time-limited to a duration of $\duration \sec$ according to
	\begin{equation}\label{eq:spill-over}
		\Ex{}{\vecnorm{\timetrunc{\duration}\inpp}^2} \geq (1-\spillover)\Ex{}{\vecnorm{\inpp}^2}
	\end{equation}
	where $0<\spillover\ll 1$; 
	\item satisfies the average-power constraint
	\begin{equation}\label{eq:av_power_constraint}
		(1/\duration)\Ex{}{\vecnorm{\inpp}^2}\leq\Pave.
	\end{equation}
\end{enumerate}

The constraints~\eqref{eq:bandwidth_constraint} and~\eqref{eq:spill-over} capture the fact that we are dealing with input signals that are strictly band-limited and essentially time-limited.
As pointed out in~\cite[p.~364]{wyner66-03a}, time limitation is important as this allows for a physically meaningful definition of transmission rate.
Note that the strict bandwidth constraint~\eqref{eq:bandwidth_constraint} implies that any nonzero~\inpp can be limited in time only  in an approximate sense~\cite{slepian76-03a}, a consideration that justifies the form of the constraint expressed in~\eqref{eq:spill-over}.

The signal~\wgnp is a zero-mean proper AWGN process with double-sided PSD equal to $1$.
Finally, the time-varying channel impulse response~\tvirp is a zero-mean jointly proper Gaussian (JPG) process in time~\time and delay~\delay that satisfies the WSSUS assumption
\begin{equation}
	\Ex{}{\tvirp \conj{\tvir}(\time',\delay')}=\corrtd(\time-\time',\delay)\dirac(\delay-\delay')
	\label{eq:WSSUS_assumption}
\end{equation}
and is independent of~\wgnp and~\inpp.
As a consequence of the JPG and the WSSUS assumptions, the \emph{time-delay correlation function}~\corrtdp fully characterizes the channel statistics.

Often, it is convenient to describe the action of the channel~$\CHop$ in  domains other than the time-delay domain used in~\eqref{eq:ltv-io}. 
Specifically, we shall frequently work  with the following alternative I/O relation [cf.~\eqref{eq:ltv-io}], which is explicit in the channel \emph{delay-Doppler spreading function}~$\spfp=\int_{\time}\tvirp  e^{-j2\pi\doppler\time}d\time$ according to
\begin{equation*}
	\outpp=\underbrace{\spreadint{\spfp\inp(\time-\delay)e^{j2\pi\doppler\time}}}_{=\,\outpnnp}+\wgnp.
\end{equation*}
This alternative I/O relation leads to the following physical interpretation: the noiseless output signal $\outpnnp=(\CHop \inp)(\time)$ is a weighted superposition of copies of the input signal~\inpp that are shifted in time by the delay~\delay and in frequency by the Doppler shift \doppler.  
The spreading function is the corresponding weighting function.   
In other words, the channel operator $\CHop$ can be represented as a continuous weighted superposition of time-frequency shift operators.
Note that every ``reasonable'' linear operator admits such a representation (see~\cite[Thm.~14.3.5]{groechenig01a} for a precise mathematical formulation of this statement). 
As a consequence of the WSSUS assumption, the spreading function~\spfp is uncorrelated in~\delay and~\doppler, i.e., we have
\begin{equation}\label{eq:scattering_function}
	\Ex{}{\spfp\conj{\spf}(\delay',\doppler')}=\scafunp\dirac(\delay-\delay')\dirac(\doppler-\doppler')
\end{equation}
where~\scafunp is the two-dimensional PSD of the channel process, usually referred to as~\emph{scattering function}~\cite{matz11-03a}.
In the remainder of the paper, we let the scattering function be normalized in volume according to
\begin{equation}
\label{eq:scafunnorm}
	\spreadint{\scafunp}=1.
\end{equation}
Another system function we shall need is the \emph{time-varying transfer function} 
\begin{equation*}	
	\tvtf(\time,\freq)\define\int_{\delay}\tvirp e^{-j2\pi \freq\delay}d\delay
\end{equation*}
which, as a consequence of~\eqref{eq:WSSUS_assumption}, is stationary in both time and frequency:
\begin{equation}
	\label{eq:2D_stationarity}	
	\Ex{}{\tvtf(\time,\freq)\conj{\tvtf}(\time',\freq')}=\corrtf(\time-\time',\freq-\freq').
\end{equation}
Here, $\corrtf(\time,\freq)$ denotes the time-frequency correlation function of the channel process, which is related to the scattering function through a two-dimensional Fourier transform
\begin{equation*}	
	\corrtf(\time,\freq)=\spreadint{\scafunp e^{j2\pi(\doppler\time-\delay\freq)}}.
\end{equation*}
For a more complete description of the WSSUS channel model, the interested reader is referred to~\cite{matz11-03a,durisi10-01a}.

\subsection{A Robust Definition of Underspread Channels} 
\label{sec:a_robust_definition_of_underspread_channels}
Qualitatively speaking, WSSUS \emph{underspread} channels are WSSUS channels with a scattering function that is highly concentrated in the delay-Doppler plane~\cite{bello63-12a}.
For the case where~\scafunp is {\it compactly supported}, the channel is said to be underspread if the support area of~\scafunp is smaller than~$1$ (see for example~\cite{kozek97-03a,durisi10-01a}).
The compact-support assumption on \scafunp, albeit mathematically convenient, is a fine detail of the channel model in the terminology introduced in \fref{sec:introduction}, because it is not possible to determine through channel measurements whether~\scafunp is compactly supported or not.  
However, the results discussed in \fref{sec:introduction}, in the context of the stationary discrete-time fading channel model, imply a high capacity sensitivity  to whether the measure of the support of the PSD is smaller than $1$ or not. 
A similar sensitivity can be expected for the continuous-time WSSUS channel model.
To quantify this sensitivity, we need to work with a more general underspread definition. 
Specifically, we replace the underspread definition based on the compact-support assumption by the following, more robust and physically meaningful, assumption: 
we say that $\CHop$ is underspread if \scafunp has a small fraction of its total volume outside a rectangle of area  much smaller than~$1$. More precisely, we have the following definition. 
\begin{dfn}\label{def:underspread}
    Let~$\maxDelay,\maxDoppler \in \positivereals, \epsilon \in[0,1]$, and let~\chsetp be the set of all Rayleigh-fading WSSUS channels~$\CHop$ with scattering function~\scafunp satisfying
	\begin{equation}
	\label{eq:underspread_definition}
		\int_{-\maxDoppler}^{\maxDoppler}\int_{-\maxDelay}^{\maxDelay} \scafunp d\delay d\doppler  \geq 1-\uncertainty.
	\end{equation}
	We say that the channels in~\chsetp are \emph{underspread} if~$\spread\define4\maxDelay\maxDoppler \ll 1$ and~$\uncertainty\ll 1$.
\end{dfn}

Note that it is possible to verify, through channel measurements, whether a fading channel is underspread according to \fref{def:underspread}.
Typical wireless channels are (highly) underspread, with most of the volume of~\scafunp supported over a rectangle of area $\spread \leq 10^{-3}$ for land-mobile channels, and~\spread as small as~$10^{-7}$ for certain indoor channels with restricted terminal mobility. 
Note that setting~$\uncertainty=0$ in~\fref{def:underspread} yields the compact-support underspread definition of~\cite{kozek97-03a,durisi10-01a}. 
The moment-based underspread definition proposed in~\cite{matz03a} is subsumed by \fref{def:underspread} as well.
\subsection{Band-Limitation at the Receiver} 
\label{sec:receive_filter}
Even though~\inpp has bandwidth no larger than~\bandwidth, the signal $\outpnnp=(\CHop \inp)(\time)$ is,  in general, not strictly band-limited, because $\CHop$ can  introduce arbitrarily large frequency dispersion. 
However, if $\CHop$ is underspread in the sense of \fref{def:underspread}, most of the energy of \outpnnp will be supported on a frequency band of size $(\bandwidth+2\maxDoppler) \Hz$.
We therefore assume that the output signal~\outpp is passed through an ideal low-pass filter of bandwidth $(\bandwidth+2\maxDoppler) \Hz$, resulting in the filtered output signal
\begin{equation}\label{eq:filt_IO}	
	\outpfiltp=(\freqtrunc{\bandwidth+2\maxDoppler}\outp)(\time).
\end{equation}
This filtering operation yields a band-limited WSSUS fading channel.

\section{Channel Capacity} 
\label{sec:the_information_capacity}%
\subsection{Outline of the Information-Theoretic Analysis} 
\label{sec:outline_of_the_information_theoretic_analysis}%
We are interested in characterizing the ultimate limit on the rate of reliable communication  over the continuous-time fading channel~\eqref{eq:ltv-io} in the noncoherent setting (i.e., the setting where neither the transmitter nor the receiver know the realization of~$\CHop$, but both know the statistics of $\CHop$).
Two main difficulties need to be overcome to obtain such a characterization.
First, we need to deal with continuous-time channels and signals, which are notoriously difficult to analyze  information-theoretically.
Second, our focus is on the noncoherent setting, for which, even for simple discrete-time channel models, analytic capacity characterizations are not available.
%
%
                                                             
To overcome these difficulties we resort to bounds on capacity.
As (trivial) capacity upper bound, we take in~\fref{sec:an_upper_bound_on_capacity}  the capacity of a band-limited Gaussian channel~\cite{wyner66-03a} with the same average-power constraint as in~\eqref{eq:av_power_constraint} and bandwidth equal to $(\bandwidth+2\maxDoppler)$.
A capacity lower bound is obtained in~\fref{sec:a_lower_bound_on_capacity} through the following  two steps: first, we  construct a discretized channel whose capacity is proven to be a lower bound on the capacity of the underlying continuous-time channel~\eqref{eq:ltv-io}; then, we derive a lower bound on the capacity of this discretized channel that is explicit in the channel parameters \spread and \uncertainty.                                
In~\fref{sec:numerical_results}, we then show that, for channels that are underspread according to~\fref{def:underspread}, this lower bound is close to the AWGN-channel capacity upper bound for all SNR values of practical interest, thereby sandwiching the capacity  of the band-limited continuous-time fading channel tightly.    
%
\subsection{Mutual Information and Capacity for the Continuous-Time Channel} 
\label{sec:mutual_information_and_capacity_in_the_continuous_time_case}
Dealing with continuous-time channels requires a suitable generalization of the definitions of mutual information and capacity~\cite{cover06-a} to the continuous-time case.  
Such a generalization can be found, e.g., in~\cite{gelfand57-a}, \cite[Ch.~8]{gallager68a}, and is reviewed here for completeness. 

To define  capacity of the channel~\eqref{eq:ltv-io}, we represent the complex signals at the input and output of~$\CHop$ in terms of projections onto complete orthonormal sets for the underlying signal spaces.
More specifically, let $\{\ofunp\}_{\oindex=0}^{\infty}$ be a complete orthonormal set for the space $\hilfunspace(\bandwidth)$ of signals with bandwidth no larger than \bandwidth.
We can then describe $\inpp \in \hilfunspace(\bandwidth)$ uniquely in terms of the projections
\begin{equation}\label{eq:input_projections}
	\inp_{\oindex}\define\inner{\inpp}{\ofunp}, \quad \oindex=0,1,\dots
\end{equation}   
as $\inpp=\sum_{\oindex}\inp_{\oindex}\ofunp$.
Similarly, let  $\{\altofunp\}_{\oindex=0}^{\infty}$ be a complete orthonormal set for $\hilfunspace(\bandwidth + 2\maxDoppler)$.  
The  low-pass filtered output signal $\outpfiltp \in \hilfunspace(\bandwidth + 2\maxDoppler)$ in~\eqref{eq:filt_IO} can be  described uniquely in terms of the projections
\begin{equation}\label{eq:output_projections}
	 \outp_{\oindex}\define\inner{\outpfiltp}{\altofunp}, \quad \oindex=0,1,\dots
\end{equation}  
as $\outpfiltp=\sum_{\oindex}\outp_{\oindex}\altofunp$.
To define the mutual information between~\inpp and~\outpfiltp, we need to impose a probability measure on~\inpp.\footnote{A probability measure on \inpp is specified through the joint probability measure of the $n$-tuples $(\inp(t_1),\dots, \inp(t_n))$ for every $n \in \naturals$ and for every choice of $t_1,\dots,t_n \in \reals$~\cite[Sec.~25.2]{lapidoth09a}.}
Concretely, let~\probinpset be the set of probability measures on~\inpp that satisfy the bandwidth constraint~\eqref{eq:bandwidth_constraint}, the time-limitation constraint~\eqref{eq:spill-over}, and the average-power constraint~\eqref{eq:av_power_constraint}.
Every probability measure in~$\probinpset$ induces a corresponding probability measure on~$\{\inp_\oindex\}_{\oindex=0}^\infty$.
For a given probability measure in $\probinpset$, the mutual information between~\inpp and~\outpfiltp is defined as~\cite[Eq.~(8.151)]{gallager68a}, \cite[Def.~3, Thm.~1.5]{gelfand57-a} 
\begin{equation*}
	\mi(\outpfiltp;\inpp)\define\lim_{\onum  \to \infty} \mi(\outpvec^{\onum};\inpvec^{\onum})
\end{equation*}    
where $\inpvec^{\onum}=\tp{[\inp_0\,\, \inp_1\, \dots\, \inp_{\onum}]}$, and, similarly, $\outpvec^{\onum}=\tp{[\outp_0\,\, \outp_1\, \dots\, \outp_{\onum}]}$. 
This definition turns out to be independent of the complete orthonormal sets $\{\ofunp\}_{\oindex=0}^{\infty}$ and $\{\altofunp\}_{\oindex=0}^{\infty}$ used~\cite[Thm.~1.5]{gelfand57-a}.  
The capacity \capacity of the channel~\eqref{eq:ltv-io} can now be defined as follows~\cite[Eq.~(8.1.55)]{gallager68a}:
\begin{equation}\label{eq:capacity_definition_ct}
	\capacity\define\lim_{\duration \to \infty} \frac{1}{\duration}\sup_{\probinpset}\mi(\outpfiltp;\inpp).
\end{equation}
We conclude this section by noting that, by Fano's inequality, no rate above \capacity is achievable~\cite{verdu94-07a}. 
However, whether the channel coding theorem applies to the general class of time-frequency selective fading channels considered in this paper is an open problem, even for the discrete-time case~\cite{koch10-12a}. 
%
%

%
%
%

\subsection{An Upper Bound on Capacity} 
\label{sec:an_upper_bound_on_capacity}

For underspread channels in~\chsetp (see \fref{def:underspread}) and input signals satisfying~\eqref{eq:bandwidth_constraint}--\eqref{eq:av_power_constraint}, we take as simple (yet tight, in a sense to be specified in \fref{sec:numerical_results}) upper bound on~\eqref{eq:capacity_definition_ct} 
 the capacity of a (nonfading) band-limited AWGN channel with the same average-power constraint as in~\eqref{eq:av_power_constraint} and bandwidth $(\bandwidth+2\maxDoppler)$.
More precisely, we show in \fref{app:awgn_capacity_upper_bound} that  $\capacity\leq\cawgn$, where         
\begin{IEEEeqnarray}{rCl}\label{eq:awgn_capacity}
	\cawgn&\define& (\bandwidth+2\maxDoppler) \log\mathopen{}\left(1+(1-\spillover)(1-\uncertainty)\frac{\Pave}{\bandwidth+2\maxDoppler}\right) \nonumber\\
	&&+ (\spillover+\uncertainty-\spillover\uncertainty)\Pave.%
\end{IEEEeqnarray}
This result is based on~\cite[Thm. 2]{wyner66-03a}.   
Differently from~\cite[Eq.~(20)]{wyner66-03a}, the second term on the right-hand side (RHS) of~\eqref{eq:awgn_capacity} accounts not only for the approximate time-limitation of \inpp, but also for the dispersive nature of $\CHop$.

It is now appropriate to provide a preview of the nature of the results we are going to obtain.  
We will show that, as long as~$\spread \ll 1$ and~$\uncertainty \ll 1$, the capacity of every channel in~\chsetp, independently of whether its scattering function is compactly supported or not, is close to the AWGN-channel capacity~\cawgn for all SNR values typically encountered in practical wireless communication systems.
To establish this result, we derive, in the next section, a lower bound on~\eqref{eq:capacity_definition_ct}. 

\section{A Lower Bound on Capacity} 
\label{sec:a_lower_bound_on_capacity} 

\subsection{Outline} 
\label{sec:outline}%
As the derivation of the capacity lower bound presented in this section consists of several steps, we start by providing an outline of our proof strategy.
The first step entails restricting the set of input distributions in~\eqref{eq:capacity_definition_ct} to a subset of~\probinpset; this clearly yields a lower bound on \capacity.
The subset of \probinpset we consider is described in~\fref{sec:a_smaller_set_of_input_distributions} and is obtained by  constraining the input signal \inpp to lie in the span of an orthonormal WH set (that is not necessarily complete for $\hilfunspace(\bandwidth)$). 
The second step (see \fref{sec:the_discretized_i_o_relation}) consists of projecting the corresponding output signal \outpfiltp onto the same orthonormal WH set, an operation that further lower-bounds mutual information, as seen by application of the  data-processing inequality~\cite[Thm.~1.4]{gelfand57-a} (the orthonormal WH set is not necessarily complete for $\hilfunspace(\bandwidth+2\maxDoppler)$).
As a result of these two steps, we obtain a discretization of the I/O relation.
The capacity of the corresponding discretized channel, which is a lower bound on the capacity of the underlying continuous-time channel, is further lower-bounded in~\fref{sec:the_lower_bound} by treating the off-diagonal terms in the I/O relation as (signal-dependent) additive noise.
This finally yields a lower bound on the capacity of the underlying continuous-time channel  that is explicit in the channel parameters \spread and \uncertainty.

\subsection{A Smaller Set of Input Distributions} 
\label{sec:a_smaller_set_of_input_distributions}
Let $\slogonp\define\logon(\time-\dtime\tstep)\cex{\dfreq\fstep\time}$ and
\begin{equation*}	
	\WHset\define\bigl\{\slogonp\bigr\}_{\dtime,\dfreq \in \integers}
\end{equation*}
be an orthonormal WH set, i.e., a set consisting of time-frequency shifts (on a rectangular lattice) of a given pulse~$\logonp \in \hilfunspacep$.  
Orthonormality of the WH set implies $\tfstep \geq 1$, as a consequence of~\cite[Cor. 7.5.1, Cor. 7.3.2]{groechenig01a}.
We lower-bound~\capacity by restricting the input signals to be of the form
\begin{equation}
\label{eq:canonical-input}
 	\inpp=\sumdtimelimited\sumdfreqlimited
 		\inp[\dtdf] \slogon(\time)
\end{equation}
where $\{\inp[\dtdf]\}$ are random coefficients. 
To guarantee that~\inpp in~\eqref{eq:canonical-input} satisfies~\eqref{eq:bandwidth_constraint}--\eqref{eq:av_power_constraint}, we impose the following constraints on \WHset, \tslots, \fslots, and $\{\inp[\dtdf]\}$.          
%

\subsubsection{Average-power constraint} 
\label{sec:the_average_power_constraint_eqref}
To ensure that~\inpp in~\eqref{eq:canonical-input}  satisfies~\eqref{eq:av_power_constraint}, it is sufficient 
to choose \tslots such that $(2\tslots+1)\tstep\leq \duration$ (further restrictions on the choice of $\tslots$ will be imposed in \fref{sec:time_limitation_constraint_eq:spill-over}), and to require that  the random variables $\{\inp[\dtdf]\}$ satisfy 
\begin{equation}
 \label{eq:apc}
  \sumdtimelimited\sumdfreqlimited
 		\Ex{}{\abs{\inp[\dtdf]}^{2}} \le (2\tslots+1)\tstep\Pave.
\end{equation}
The constraint~\eqref{eq:apc}, together with the orthonormality of the set \WHset, implies that~\eqref{eq:av_power_constraint} is satisfied.

\subsubsection{Bandwidth limitation} 
\label{sec:bandwidth_constraint_eq:bandwidth_constraint}
  To ensure that~\inpp in~\eqref{eq:canonical-input}  satisfies~\eqref{eq:bandwidth_constraint}, we require that~\logonp fulfills the following property.
\begin{prop}\label{prop:bandwidth}
	 The function \logonp is strictly band-limited with bandwidth $\fstep\leq \bandwidth$.
\end{prop} 
                                                                              
Furthermore, we take $\fslots=(\fslotstot-1)/2$ where $\fslotstot\define\bandwidth/\fstep$. 
For simplicity of exposition, we shall assume, in the remainder of the paper, that $\fslotstot$ is an odd  integer.

\subsubsection{Time limitation} 
\label{sec:time_limitation_constraint_eq:spill-over}
To ensure that $\inpp$ in~\eqref{eq:canonical-input} satisfies~\eqref{eq:spill-over}, we impose two additional constraints.
First, we require that $\logonp$ satisfies the following property.  
\begin{prop}\label{prop:decay}
	The function \logonp is even and decays faster than $1/\time$, i.e., 
	\begin{equation}\label{eq:decay}
		\logonp=\landauO(1/\time^{1+\decay}),\quad \time \to \infty 
	\end{equation}
	for some $\decay>0$.
\end{prop}  
\begin{figure}[t]
	\centering
		\includegraphics[width=0.9\figwidth]{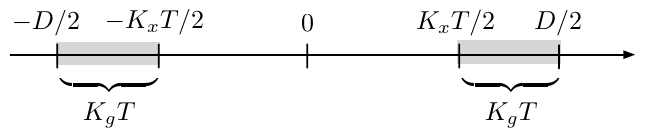}
	\caption{Insertion of guard intervals.}
	\label{fig:guard_intervals}
\end{figure}

Second, we insert, in the interval $[-\duration/2,\duration/2]$, two \emph{guard intervals}.
More specifically,  for a given approximate duration \duration of the input signal \inpp [we will later take 
$\duration \to \infty$ according to~\eqref{eq:capacity_definition_ct}], the interval $[-\duration/2,\duration/2]$ is divided up into three parts (see \fref{fig:guard_intervals}): the interval $[-\tslotstot\tstep/2,\tslotstot\tstep/2]$, with $\tslots=(\tslotstot-1)/2$ in~\eqref{eq:canonical-input},\footnote{We assume that $\tslotstot$ is an odd integer.} supporting most of the energy of \inpp, and two guard intervals $[-\duration/2,-\tslotstot\tstep/2]$ and $[\tslotstot\tstep/2,\duration/2]$, each of length $\tslotsguard\tstep=\duration/2-\tslotstot\tstep/2$.
This will ensure that~\eqref{eq:spill-over} is satisfied.
We will let $\tslotstot \to \infty$ as $\duration \to \infty$, with $\tslotsguard$ kept constant.
This guarantees that  the fraction of time allocated to the guard intervals vanishes as $\duration \to \infty$. 
For simplicity of notation, we shall assume in the remainder of the paper that $\tslotsguard$ is an integer. 
For fixed \spillover in~\eqref{eq:spill-over}, the decay property of $\logonp$ expressed in~\eqref{eq:decay}  implies that one can choose $\tslotsguard$ (independent of \tslots) so that \inpp in~\eqref{eq:canonical-input} satisfies~\eqref{eq:spill-over}.
This statement is proven in~\fref{app:proof_of_lemwh_and_spillover}. 

We next show formally that our construction results in a capacity lower bound.
Fix an orthonormal WH set \WHset satisfying Properties~\ref{prop:bandwidth} and \ref{prop:decay}.
Furthermore, let $\probinpdisc$ be the set of probability measures on $\{\inp[\dtdf]\}$ that satisfy~\eqref{eq:apc}. 
Every probability measure in $\probinpdisc$ induces a probability measure on \inpp in~\eqref{eq:canonical-input}.
We denote the corresponding set of probability measures on \inpp by \probinpwhset.
As just shown, \inpp satisfies~\eqref{eq:bandwidth_constraint}--\eqref{eq:av_power_constraint}.
Hence, $\probinpwhset \subseteq \probinpset$ [recall that \probinpset is the set of \emph{all} probability measures that satisfy \eqref{eq:bandwidth_constraint}--\eqref{eq:av_power_constraint}].
We can then lower-bound \capacity in~\eqref{eq:capacity_definition_ct} as follows:
\begin{align}\label{eq:capacity_lb1} 
	  \capacity &=\lim_{\duration \to \infty} \frac{1}{\duration}\sup_{\probinpset}\mi(\outpfiltp;\inpp) \notag\\   
	   &\geq  \lim_{\duration \to \infty} \frac{1}{\duration}\sup_{\probinpwhset}\mi(\outpfiltp;\inpp).
\end{align}
Here, the inequality follows by restricting the supremization to the smaller set~\probinpwhset.

\subsection{The Discretized I/O Relation} 
\label{sec:the_discretized_i_o_relation}
The second step in our approach is to project the output signal~\outpfiltp [resulting from the transmission of \inpp in~\eqref{eq:canonical-input}] onto
the signal set $\left\{g_{k,n}(t)\right\}$ to obtain 
\begin{IEEEeqnarray}{rCl}
	\outp[\dtdf]&\define& \inner{\outpfilt}{\slogon} \notag\\
	&\stackrel{(a)}{=}& \inner{\outp}{\slogon} 
	\notag\\
	&=&\underbrace{\inner{\CHop\slogon}{\slogon}}_{\define\,\ch[\dtdf]} \inp[\dtdf]\notag\\
	&&+\mathop{\altsumdtimelimited\altsumdfreqlimited}_{(\altdtdf)\neq(\dtdf)} \underbrace{\inner{\CHop\logon_{\altdtdf}}{\slogon}}_{\define\,\interfp}\inp[\altdtdf] +\underbrace{\inner{\wgn}{\slogon}}_{\define\,\wgn[\dtdf]}\notag\\
	&=&\ch[\dtdf]\inp[\dtdf] \notag\\
	&&+\mathop{\altsumdtimelimited\altsumdfreqlimited}_{(\altdtdf)\neq(\dtdf)} \interfp\inp[\altdtdf]
	+\wgn[\dtdf]
	\label{eq:discretized I/O with interference}
\end{IEEEeqnarray}
for each \emph{time-frequency} slot~$(\dtime,\dfreq)$,
$\dtime=-\tslots, -\tslots+1, \dots, \tslots$, $\dfreq = -\fslots, -\fslots+1, \dots, \fslots$.
Here, (a) is a consequence of~\fref{prop:bandwidth}, which implies that the Fourier transform of \slogonp (with $\dtime=-\tslots, -\tslots+1, \dots, \tslots$, $\dfreq = -\fslots, -\fslots+1, \dots, \fslots$) is strictly supported in the interval $[-\bandwidth/2,\bandwidth/2]$.  
We refer to the channel with I/O relation~\eqref{eq:discretized I/O with interference} as the discretized channel \emph{induced} by the  WH set~\WHset. 
As we assumed that $\tvirp$ in~\eqref{eq:ltv-io} is a zero-mean JPG random process in \time and \delay, the random variables $\ch[\dtdf]$ and $\interfp$ are zero-mean JPG.   
Furthermore, the orthonormality of the WH set~\WHset implies that the~$\wgn[\dtdf]$ in~\eqref{eq:discretized I/O with interference} are \iid $\jpg(0,1)$.    
For each time slot~$\dtime \in \{-\tslots,-\tslots+1,\ldots,\tslots\}$, we arrange the data  symbols~$\inp[\dtdf]$, the output signal samples~$\outp[\dtdf]$, the channel coefficients~$\ch[\dtdf]$, and the noise samples~$\wgn[\dtdf]$ in corresponding \fslotstot-dimensional vectors.\footnote{Recall that $\tslotstot=2\tslots+1$ and $\fslotstot=2\fslots+1$.}
For example, the \fslotstot-dimensional vector that contains the input symbols in the $\dtime$th time slot is defined as
\begin{align*}
		\inpvec[\dtime]&\define\tp{\mat\inp[\dtime,-\fslots]\; \inp[\dtime,-\fslots+1]\; \ldots\;
			\inp[\dtime,\fslots]\emat}.
\end{align*}
The output vector $\outpvec[\dtime]$, the channel vector $\chvec[\dtime]$, and the noise vector $\wgnvec[\dtime]$ are defined analogously.
To get a compact notation, we further stack~$\tslotstot$ contiguous  input, output, channel, and noise vectors, into  corresponding $\tslotstot\fslotstot$-dimensional vectors.
For example, for the channel input this results in the $\tslotstot\fslotstot$-dimensional vector
\begin{align}
	\inpvec&\define\tp{\mat\tp{\inpvec}[-\tslots]\;\; \tp{\inpvec}[-\tslots+1]\;\; \ldots\;\;
			\tp{\inpvec}[\tslots]\emat}\label{eq:mv_input}.
\end{align}
Again, the stacked vectors \outpvec, \chvec, and \wgnvec are defined analogously.
Finally, we arrange the self-interference terms~\interfp  in a~$\tslotstot\fslotstot \times \tslotstot\fslotstot$ matrix~\interfmat with entries
\begin{multline*}
[\interfmat]_{\dfreq+\dtime\fslotstot,\altdfreq+\altdtime\fslotstot}\\
=\begin{cases} 
\interf[\altdtime-\tslots,\altdfreq-\fslots,\dtime-\tslots,\dfreq-\fslots],  &\text{if }  (\altdtdf) \neq (\dtdf) \\
0, &\text{otherwise}
\end{cases}
\end{multline*}
for $\allz{\altdtime,\dtime}{\tslotstot}$
and $\allz{\altdfreq,\dfreq}{\fslotstot}$.
With these definitions, we can now compactly express the I/O relation~\eqref{eq:discretized I/O with interference} as
\begin{equation}
\label{eq:I_O_in_vector_matrix_form}
	\outpvec=\chvec\had\inpvec+ \interfmat\inpvec +\wgnvec.
\end{equation}

Let now \capacitydisc be the capacity of the discretized channel~\eqref{eq:I_O_in_vector_matrix_form} [induced by the WH set \WHset\!] with \inpvec subject to the average-power constraint~\eqref{eq:apc}.  
We can lower-bound the RHS of~\eqref{eq:capacity_lb1} by \capacitydisc as follows
\begin{align}\label{eq:capacity_disc}
	 \capacity  &\stackrel{(a)}{\geq}  \lim_{\duration \to \infty} \frac{1}{\duration}\sup_{\probinpwhset}\mi(\outpfiltp;\inpp) \notag\\ 
	&\stackrel{(b)}{\geq}\lim_{\tslotstot \to \infty} \frac{1}{(\tslotstot+2\tslotsguard)\tstep}\sup_{\probinpdisc} \mi(\outpvec;\inpvec)\notag\\
	&\define    \capacitydisc.
\end{align}
Here, in~(a) we used~\eqref{eq:capacity_lb1}, and~(b) is a consequence of~\cite[Thm. 1.4]{gelfand57-a}, which extends the data processing inequality to continuous-time signals.   
To summarize, we showed that the capacity of the discretized channel~\eqref{eq:I_O_in_vector_matrix_form} induced by the WH set \WHset is a lower bound on the capacity of the underlying continuous-time channel~\eqref{eq:ltv-io}.    
%
%
                                 
\subsection{Why Weyl-Heisenberg Sets?} 
\label{sec:why_wh_sets_}
The choice of constraining \inpp to lie in the span of an orthonormal WH set according to~\eqref{eq:canonical-input} results in a signaling scheme that can be interpreted as PS-OFDM~\cite{kozek98-10a}, where the data symbols 
$\inp[\dtdf]$ are modulated onto a set of orthogonal signals indexed by discrete time (symbol index) \dtime, and discrete frequency (subcarrier index) \dfreq.
From this perspective, the self-interference term (the second term on the RHS of~\eqref{eq:discretized I/O with interference}, which is made up of the off-diagonal terms in the I/O relation) can be interpreted as intersymbol and intercarrier interference.
Discretization through WH sets is sensible for the following two reasons.
\paragraph*{Stationarity} 
\label{par:stationarity}
The structure of WH sets preserves the stationarity of the channel in the discretization.
More precisely, the channel gains $\ch[\dtdf]$ in~\eqref{eq:discretized I/O with interference}
inherit the two-dimensional stationarity property of the underlying continuous-time channel [see~\eqref{eq:2D_stationarity}], a fact that is crucial for the ensuing analysis.
We prove this result in \fref{app:statistical_properties}, where we also establish properties of the statistics of~$\interfp$ in~\eqref{eq:discretized I/O with interference} that will be needed in the remainder of the paper.

\paragraph*{Approximate diagonalization} 
\label{par:approximate_diagonalization}
The presence of the self-interference term in~\eqref{eq:discretized I/O with interference} makes the computation of \capacitydisc in~\eqref{eq:capacity_disc} involved.
A classic approach to eliminate self-interference is to discretize the channel by projecting the input and output signals onto the channel-operator singular functions~\cite{wyner66-03a,gallager68a}.
This choice is convenient, as it leads to a diagonal discretized I/O relation, i.e., to countably many scalar, non-interacting I/O relations (see~\cite{durisi11-03b} for more details).
Unfortunately, this approach is not viable in our setup, because in the LTV case the channel-operator singular functions are, in general, random and not known to transmitter and receiver (recall that we consider the noncoherent setting). 
Discretizing using \emph{deterministic} orthonormal functions, as done in the previous section, yields self-interference, which we will need to take into account.
This will be accomplished by treating self-interference as additive noise, which will further lower-bound capacity. 
The main technical difficulty in this context arises from the self-interference term being signal-dependent. 
Moreover, as our capacity lower bound is obtained by treating self-interference as noise, ensuring that the power in the self-interference term is small (and, hence, that the discretized I/O relation is approximately diagonal) is crucial to get a good capacity lower bound.
This can be accomplished by choosing the pulse \logonp to be well localized in time and frequency.
In fact, it was shown in~\cite{kozek97-03a,matz98-08a,matz03a,durisi10-01a} that the singular functions of random underspread operators can be well approximated by  orthonormal WH sets generated by pulses that are well localized in time and frequency.

\subsection{A Lower Bound on the Capacity of the Discretized Channel} 
\label{sec:the_lower_bound} 
We next derive a lower bound on \capacitydisc [and, hence, on~\capacity in~\eqref{eq:capacity_definition_ct}] by using a Gaussian input distribution, and by  treating self-interference as (signal-dependent) noise.
This lower bound---evaluated for an appropriately chosen WH set---will then be shown to be close (for all SNR values of practical interest) to the AWGN-channel capacity upper bound~\cawgn in~\eqref{eq:awgn_capacity}, whenever the channel is underspread according to~\fref{def:underspread}, thereby sandwiching the capacity of the underlying continuous-time channel tightly. 

Our first result is a lower bound on \capacitydisc, which we indicate as \LB, that is explicit in the 
power spectral density  \chspecfunmatp of the multivariate stationary channel process~$\{\chvec[\dtime]\}$ with autocorrelation function $\mvchcovmat[\dtime'-\dtime]\define\Ex{}{\chvec[\dtime']\herm{\chvec}[\dtime]}$, where
\begin{align}
		\chspecfunmatp\define\sum_{\dtime=-\infty}^{\infty}		\mvchcovmat[\dtime]\cexn{\dtime\specparam},\quad\abs{\specparam}\le\frac{1}{2}.
	\label{eq:mvspec}
\end{align}
We then show in~\fref{cor:LB}, \fref{sec:a_less_tight_lower_bound} that \LB can be further lower-bounded by an expression that is explicit in the channel parameters \spread and \uncertainty introduced in \fref{def:underspread}.
 
\begin{thm}\label{thm:lower bound}
Let~\WHset  be an orthonormal WH set satisfying Properties~\ref{prop:bandwidth} and~\ref{prop:decay} in \fref{sec:a_smaller_set_of_input_distributions} and consider a Rayleigh-fading WSSUS channel (not necessarily underspread)
with scattering function~$\scafunp$.
For a given  bandwidth~\bandwidth and a given SNR~$\SNR \define\Pave/\bandwidth$,  the capacity of the discretized channel~\eqref{eq:I_O_in_vector_matrix_form} induced by~\WHset is lower-bounded according to $\capacitydisc(\SNR)\geq \LB(\SNR)$, where
\begin{IEEEeqnarray}{rCl}
\label{eq:lower bound explicit in scattering function}
\LB(\SNR) &=& \frac{\bandwidth}{\tfstep} \Ex{\ch}{	\log\mathopen{}\left( 1 + \frac{\chcorr[0,0]\tfstep\SNR\abs{\ch}^{2}}{1 + \tfstep\SNR\,\intvartot } \right)} \notag\\
&&- \inf_{0<\noisesplit<1}\Biggl\{ \frac{1}{\tstep}\intdiscrete \logdet{\matI + \frac{\tfstep\SNR}{\noisesplit}\chspecfunmatp} d\specparam \notag\\
&&+\:\frac{\bandwidth}{\tfstep} \log\mathopen{}\left(1 + \frac{\tfstep\SNR}{1-\noisesplit}\intvartot\right) \Biggr\}.
\end{IEEEeqnarray}
Here,
\begin{align*}
	&\ch\distas\jpg(0,1)\nonumber\\
	&\chcorr[0,0]\define\displaystyle\spreadint{\scafunp \abs{{\af}_{\logon}(\dd)}^{2}}\nonumber\\
	&\intvartot
	\define\mathop{\sum_{\dtime=-\infty}^{\infty}\sum_{\dfreq=-\infty}^{\infty}}_{(\dtdf)\neq(0,0)} \spreadint{\scafunp \abs{\af_{\logon}(\delay -\dtime\tstep,\doppler -\dfreq\fstep)  }^{2}}
\end{align*}
where \afp denotes the ambiguity function of \logonp~(see \fref{app:statistical_properties})
and~$\chspecfunmatp$, defined in~\eqref{eq:mvspec},  denotes the matrix-valued power spectral density
of the discretized channel induced by~\WHset. 
\end{thm}
\begin{IEEEproof}
	See \fref{app:proof_lb}.
\end{IEEEproof}

\subsection{A Lower Bound that is Explicit in the Channel Parameters \spread and \uncertainty} 
\label{sec:a_less_tight_lower_bound}
For the purposes of our analysis, it is convenient to further lower-bound~\LB to get an expression that is explicit in the channel parameters \spread and \uncertainty introduced in \fref{def:underspread}.  
The resulting lower bound, presented in the next corollary, will allow us to assess how sensitive capacity is to whether \scafunp is compactly supported or not. 

\begin{cor}
\label{cor:LB}
Let~\WHset  be an orthonormal WH set satisfying Properties~\ref{prop:bandwidth} and \ref{prop:decay} in \fref{sec:a_smaller_set_of_input_distributions} and consider a  Rayleigh-fading WSSUS channel (not necessarily underspread) in the set~\chsetp
with  scattering function~$\scafunp$.
For a given  bandwidth~\bandwidth and a given SNR~$\SNR=\Pave/\bandwidth$, and under the technical condition~$\altspread\define2\maxDoppler\tstep< 1$, the capacity of the discretized channel~\eqref{eq:I_O_in_vector_matrix_form} induced by~\WHset is lower-bounded as $\capacitydisc(\SNR) \ge\LBsimple(\SNR)$, where
	\begin{IEEEeqnarray}{rCl}
	\label{eq:capacity_lower_bound_explicit_in_spread_and_uncertainty}
	\LBsimple(\SNR)
	&\define&\frac{\bandwidth}{\tfstep}\Biggl\{ \Ex{\ch}{	\log\mathopen{}\left( 1 + \frac{\tfstep\SNR(1-\uncertainty) \minambsquare\! \abs{\ch}^{2}}{1 + \tfstep\SNR(\maxsumambsquare+\uncertainty) } \right)} \notag\\
	&&- \inf_{0<\noisesplit<1}\Biggl[ \altspread \log\mathopen{}\left(1 + \frac{\tfstep\SNR}{\noisesplit\altspread}\right) \notag \\
	&&+\: (1-\altspread)\log\mathopen{}\left(1 +\frac{\tfstep\SNR\,\epsilon}{\noisesplit(1-\altspread)}\right)\notag \\
	&&+ \log\mathopen{}\left(1 +\frac{\tfstep\SNR}{1-\noisesplit}(\maxsumambsquare +\uncertainty) \right)\Biggr]\Biggr\}.
	\end{IEEEeqnarray}
	Here,~$\ch\distas\jpg(0,1)$, $\displaystyle	\minambsquare\define\min_{(\dd) \in \spreadset} \abs{\afp}^2$, and 
	\begin{equation*}
	\maxsumambsquare\define\max_{(\dd) \in \spreadset} \mathop{\sumdtime\sumdfreq}_{(\dtdf)\neq(0,0)}\abs{\af_{\logon}(\delay -\dtime\tstep,\doppler -\dfreq\fstep)  }^{2}
	\end{equation*}
	with $\spreadset \define [-\maxDelay,\maxDelay] \times [-\maxDoppler,\maxDoppler]$.
\end{cor}
\begin{IEEEproof}
	See \fref{app:a_lower_bound_explicit_in_spread_and_uncertainty}.
\end{IEEEproof}

	The lower bound~\LBsimple in~\eqref{eq:capacity_lower_bound_explicit_in_spread_and_uncertainty}  depends on the seven quantities $(\SNR,\logonp,\tstep,\fstep,\maxDelay,\maxDoppler,\uncertainty)$ and is therefore difficult to analyze. 
	We show next that if~\tstep and~\fstep are chosen so that~$\maxDoppler\tstep=\maxDelay\fstep$, a condition often referred to as the \emph{grid matching rule}~\cite[Eq. (2.75)]{kozek97-03a}, two of these seven quantities can be dropped without loss of generality. 
\begin{lem}\label{lem:square_setting}
	Let~\WHset be an orthonormal WH set satisfying Properties \ref{prop:bandwidth} and \ref{prop:decay} in \fref{sec:a_smaller_set_of_input_distributions}. Then, for any~$\beta>0$, we have
	\begin{multline*}
	\LBsimple(\SNR,\logonp,\tstep,\fstep,\maxDelay,\maxDoppler,\uncertainty)\\
	=\LBsimple\mathopen{}\left(\SNR,\sqrt{\beta}\logon(\beta \time),\frac{\tstep}{\beta},\beta\fstep,\frac{\maxDelay}{\beta},\beta\maxDoppler,\uncertainty\right).
	\end{multline*}
	In particular, assume that~$\maxDoppler\tstep=\maxDelay\fstep$ and let~$\beta=\sqrt{\tstep/\fstep}=\sqrt{\maxDelay/\maxDoppler}$ and~$\altlogonp=\sqrt{\beta}\logon(\beta \time)$. 
	Then,
	\begin{IEEEeqnarray}{rCl}	
	\IEEEeqnarraymulticol{3}{l}{ 
	\LBsimplep  
	} \nonumber \\ 
	 \quad &=&\LBsimple\mathopen{}\left(\SNR,\altlogonp,\sqrt{\tfstep},\sqrt{\tfstep},\sqrt{\spread}/2,\sqrt{\spread}/2,\uncertainty\right) \notag\\
\quad &\define&\LBsquarep.\label{eq:LB_trasformation}
	\end{IEEEeqnarray}
\end{lem}
\begin{IEEEproof}
 See \fref{app:proof_of_lem:square_setting}.
\end{IEEEproof}

In~\eqref{eq:LB_trasformation}, the superscript $(s)$ indicates that the scattering function is supported on a square (with sidelength $\sqrt{\spread}$).
In the remainder of the paper, for the sake of simplicity of exposition, we will choose~\tstep and~\fstep such that the grid matching rule~$\maxDoppler\tstep=\maxDelay\fstep$ is satisfied. 
Then, as a consequence of \fref{lem:square_setting}, we can (and will) only consider  WH sets of the form~\WHsetsquare and WSSUS channels in the set~$\chset(\sqrt{\spread}/2,\sqrt{\spread}/2,\uncertainty)$.

The lower bound \LBsquare in~\eqref{eq:LB_trasformation}  can be tightened by maximizing it over all  WH sets \WHsetsquare satisfying Properties~\ref{prop:bandwidth} and~\ref{prop:decay} in \fref{sec:a_smaller_set_of_input_distributions}. 
This maximization implicitly provides an information-theoretic criterion for choosing~\logonp and \tfstep.
Unfortunately, an analytic maximization of~\LBsquare seems complicated as the dependency of~\minambsquare and \maxsumambsquare on~\WHsetsquare is difficult to characterize analytically. 
We shall therefore choose a specific \logonp, detailed in the next section, and numerically maximize~\LBsquare as a function of \tfstep.

\subsection{A Simple WH Set} 
\label{sec:a_simple_pulse}
We  next construct a family of WH  sets~$\WHsetsquare$ that satisfy Properties~\ref{prop:bandwidth} and~\ref{prop:decay} in \fref{sec:a_smaller_set_of_input_distributions}, and has~\logonp real-valued.
Take~$1<\tfstep<2$, let~$\dur\define\srtfstep$,~$\rolloff\define\tfstep-1$, and~$\logonfp\define\four\{\logonp\}$. 
We choose~\logonfp as the (positive) square root of a raised-cosine pulse:
\begin{equation}
\label{eq:raised_cosine}
\logonf(\freq)=
\begin{cases}
\sqrt{\dur}, 
& \text{if}\quad\abs{\freq}\leq \frac{1-\rolloff}{2\dur} \\
\sqrt{\frac{\dur}{2}  (1 +\auxfun(\freq))},
& \text{if}\quad\frac{1-\rolloff}{2\dur} \leq \abs{\freq}\leq \frac{1+\rolloff}{2\dur} 
 \\
0, & \text{otherwise}
\end{cases}
\end{equation}
where~$\auxfun(\freq)\define\cos\mathopen{}\left[ \frac{\pi \dur}{\rolloff}\left(\abs{\freq}-\frac{1-\rolloff}{2\dur}\right)\right]$.
As~${(1+\rolloff)}/(2\dur)=
{\dur}/{2}$, the function~\logonfp is supported on an interval of length $\dur=\srtfstep$.
Furthermore,~\logonfp has unit norm, is real-valued and even, and  satisfies 
\begin{equation*}
	\sumdfreq \logonf(\freq-\dfreq/\dur)\logonf(\freq-\dfreq/\dur-\dtime\dur)=\dur\krond[\dtime].
\end{equation*}
By~\cite[Thm. 8.7.2]{christensen03a}, we can therefore conclude that the WH set~$(\logonp,1/\srtfstep,1/\srtfstep)$  is a tight WH frame for~\hilfunspacep, and, by duality~\cite{daubechies95-a,janssen95-a,ron97-a}, the WH set~$(\logonp,\srtfstep,\srtfstep)$ is orthonormal. 
Finally, it can be shown  that $\logonp=\landauO(1/\time^2)$ whenever $\tfstep>1$. 

\section{Finite-SNR Analysis of the Lower Bound~\LBsquare} 
\label{sec:numerical_results}
We now study  the behavior of the lower bound~\LBsquare  in~\eqref{eq:LB_trasformation} evaluated for the WH set constructed in the previous section, under the assumption that the underlying channel is underspread according to~\fref{def:underspread}, i.e., $\spread \ll1$ and~$\uncertainty \ll 1$. 
Specifically, we compare~\LBsquare to the upper bound \cawgn in~\eqref{eq:awgn_capacity}.
To simplify the comparison, we assume throughout this section that $\bandwidth\gg \maxDoppler$ (a reasonable assumption for most wireless communication systems of practical interest).
Furthermore, in~\eqref{eq:spill-over} we take  $\spillover\ll1$.
Under these assumptions, we have
\begin{equation}\label{eq:awgn_cap_approx}	
	\cawgn(\snr)\approx \bandwidth \bigl[\log\mathopen{}\left(1+(1-\uncertainty)\snr \right) + \uncertainty \snr\bigr].
\end{equation}
%
\subsection{Trade-off between Self-Interference and Signal-Space Dimensions}
\begin{figure}
	\centering
	\subfigure[]{\includegraphics[width=\figwidth]{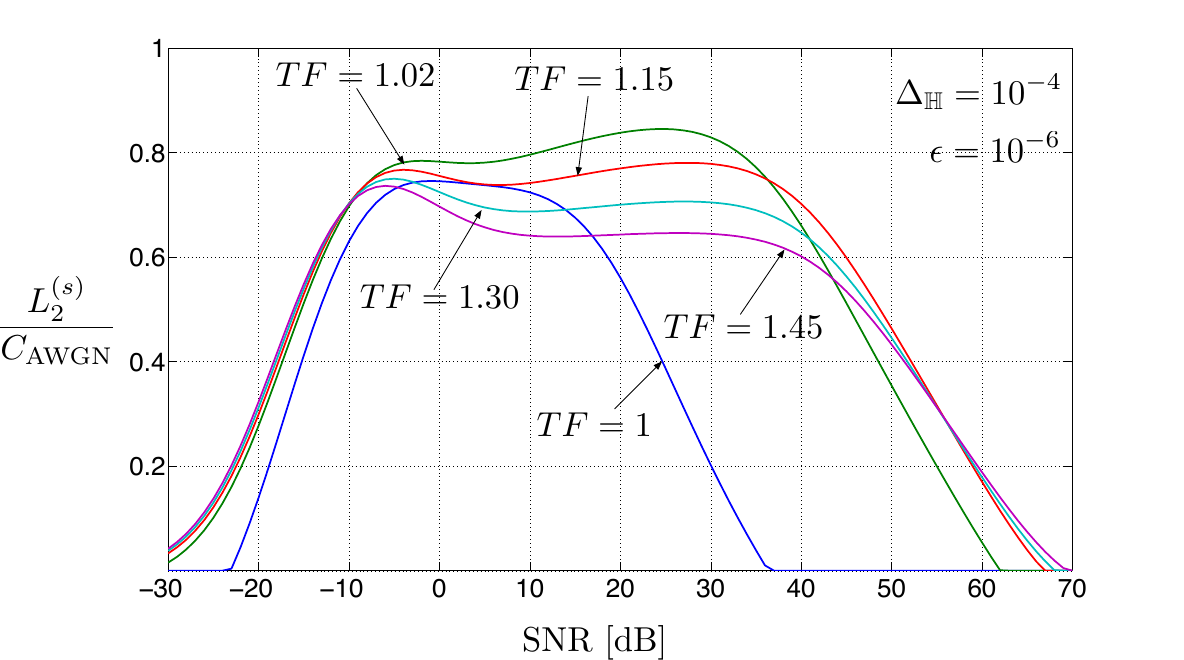}}\\
	\subfigure[]{\includegraphics[width=\figwidth]{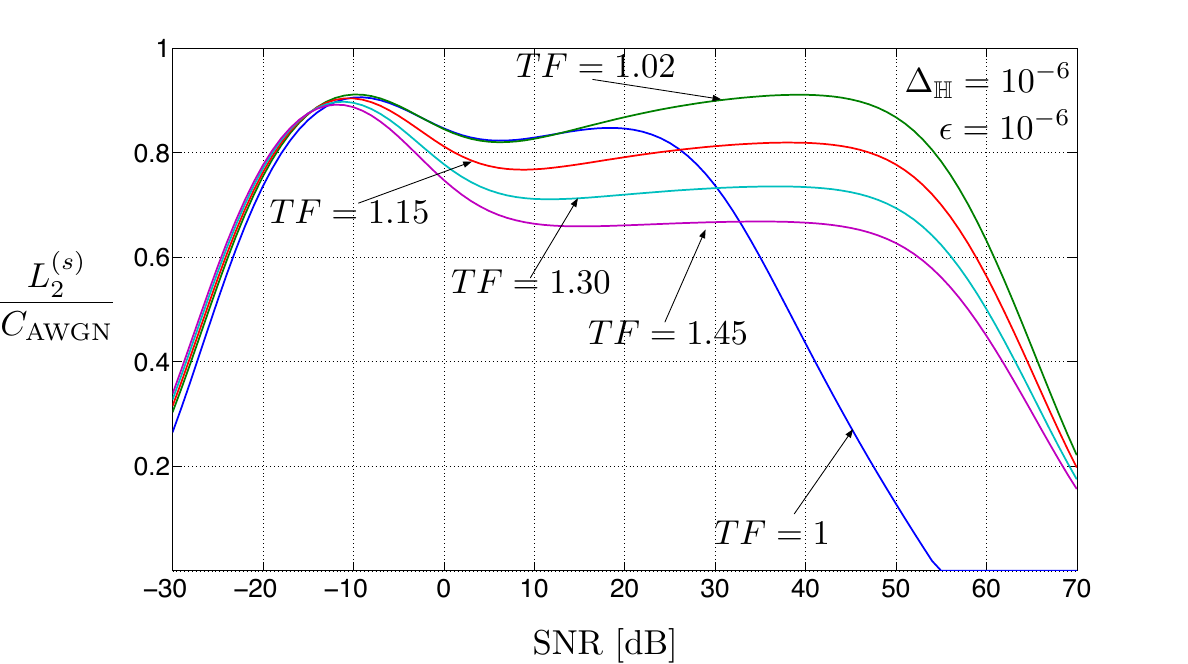}}
	\caption{Lower bounds~\LBsquare normalized with respect to the upper bound \cawgn. 
	The  bounds are computed for  WH sets based on the root-raised-cosine pulse~\eqref{eq:raised_cosine}, for different values of the grid-parameter product~\tfstep. 
	$\spread=10^{-4}$ in  (a) and  $\spread=10^{-6}$ in (b).
	In both cases, $\uncertainty=10^{-6}$.}
	\label{fig:lbnormspread4and6}
\end{figure}
In~\fref{fig:lbnormspread4and6}, we plot~$\LBsquare/\cawgn$ 
 for $\spread =10^{-4}$  and for $\spread=10^{-6}$. 
In both cases, we take  $\uncertainty =10^{-6}$.
The different curves correspond to different values of~\tfstep.   
We observe that the choice $\tfstep=1$ is highly suboptimal. 
The reason for this suboptimality is the poor time-frequency localization of~\logonp this choice entails. 
In fact, when~$\tfstep=1$, the pulse~\logonp reduces to a~$(\sin t)/t$ function, which has poor time localization.  
This, in turn, yields an ambiguity function \afp that is poorly localized in \delay, and, hence to a small value for \minambsquare and a large value for \maxsumambsquare, i.e., to small \emph{signal-to-interference ratio} (SIR) $\minambsquare/\maxsumambsquare$; this leads to a loose lower bound \LBsquare (recall that \LBsquare was obtained by treating self-interference as noise).    
\begin{figure}[t]
	\centering
		\includegraphics[width=\figwidth]{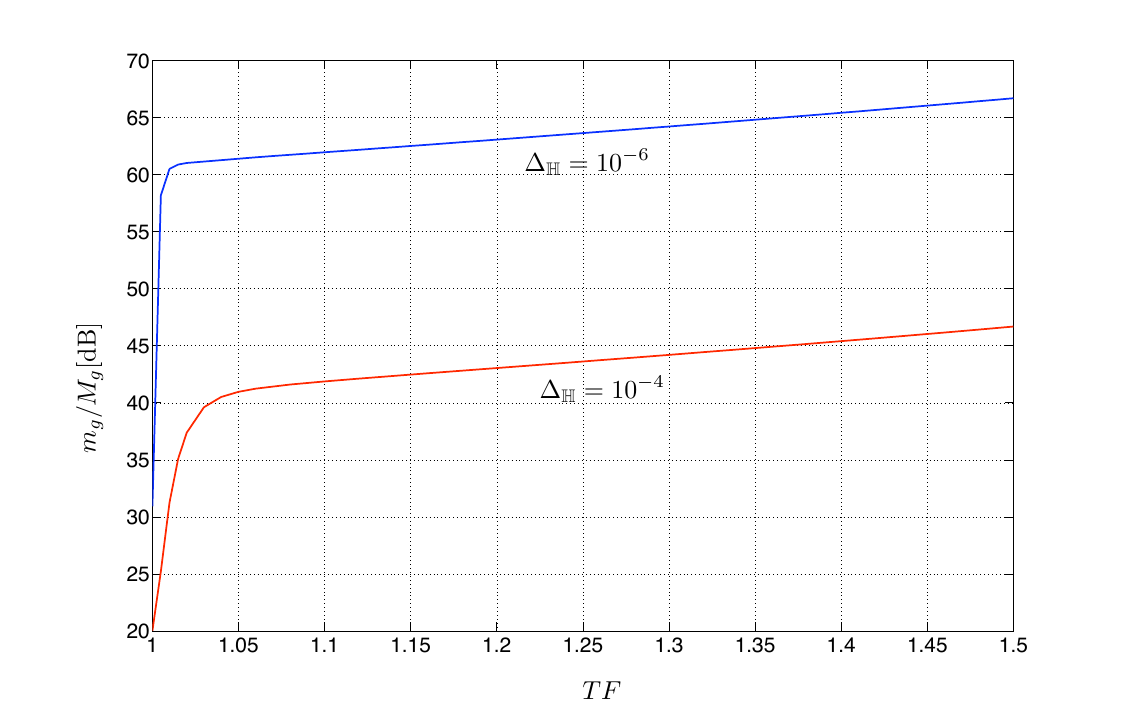}
	\caption{Trade-off between the product \tfstep, and the signal-to-interference ratio $\minambsquare/\maxsumambsquare$ for the root-raised-cosine WH set constructed in \fref{sec:a_simple_pulse}.}
	\label{fig:SIR}
\end{figure}
A value of~$\tfstep$ slightly larger than $1$ results in a significant improvement in the SIR $\minambsquare/\maxsumambsquare$ (see \fref{fig:SIR}), which is caused by the improved time localization of~\logonp.
This, in turn, yields an improved lower bound~\LBsquare for all SNR values of practical interest, as shown in \fref{fig:lbnormspread4and6}.
A further increase of the product~$\tfstep$ seems to be detrimental for all but very high SNR values, where the ratio $\LBsquare/\cawgn$ is much smaller than $1$ anyways. 
The reason underlying this behavior is as follows: in the regime where \LBsquare is close to \cawgn, the first term on the RHS of~\eqref{eq:capacity_lower_bound_explicit_in_spread_and_uncertainty} dominates the other terms. 
But in this regime, the first term on the RHS of~\eqref{eq:capacity_lower_bound_explicit_in_spread_and_uncertainty} is essentially linear\footnote{Recall that $\snr=\Pave/\bandwidth$.} in $\bandwidth/(\tfstep)$, which can be interpreted as the number of signal-space dimensions available for communication.
The loss of signal-space dimensions incurred by choosing $\tfstep$  much larger than $1$ quickly outweighs the SIR gain resulting from improved time-frequency localization. 
Our numerical results suggest that  a value of~\tfstep slightly larger than $1$ optimally trades signal-space dimensions for SIR maximization. 
We hasten to add that this trade-off is a consequence of self-interference being treated as (signal-dependent) noise in deriving our lower bound.

\subsection{Sensitivity of Capacity to the Channel Parameters~\spread and~\uncertainty}\label{sec:numerical_sensitivity}
\begin{figure}
	\centering
		\includegraphics[width=\figwidth]{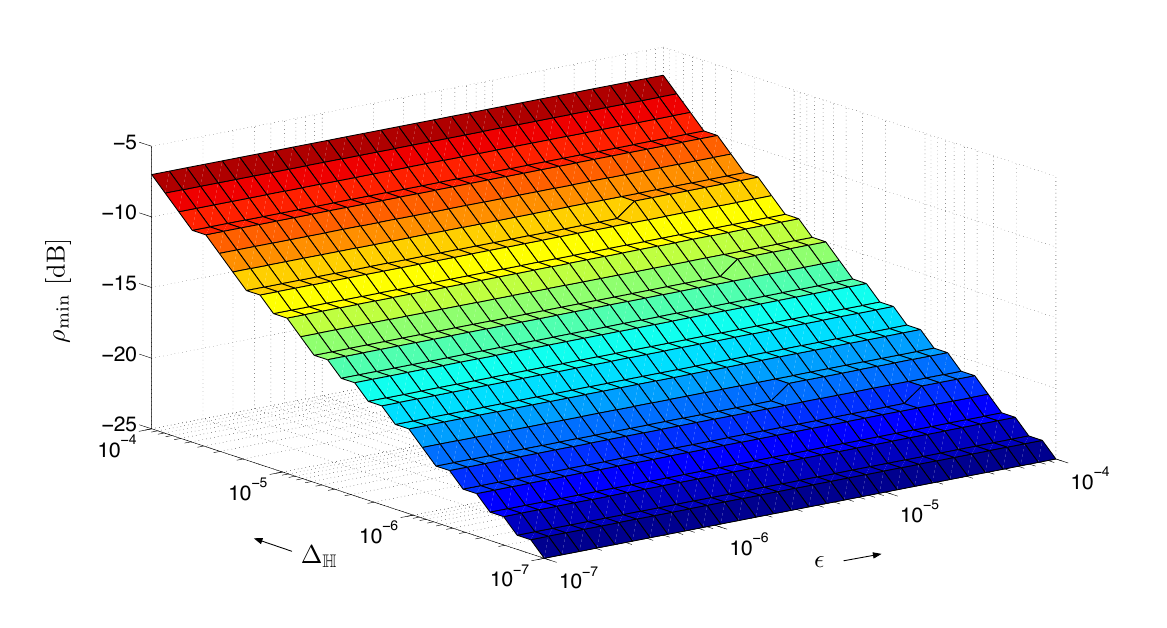}
	\caption{Minimum SNR value~\SNRmin for which~\eqref{eq:accuracy_inequality} holds, as a function of~\spread and~\uncertainty.
		The lower bound~\LBsquare is evaluated for a WH set based on the root-raised-cosine pulse~\eqref{eq:raised_cosine}; furthermore, $\tfstep=1.02$.}
	\label{fig:min}
\end{figure}

\begin{figure}
	\centering
		\includegraphics[width=\figwidth]{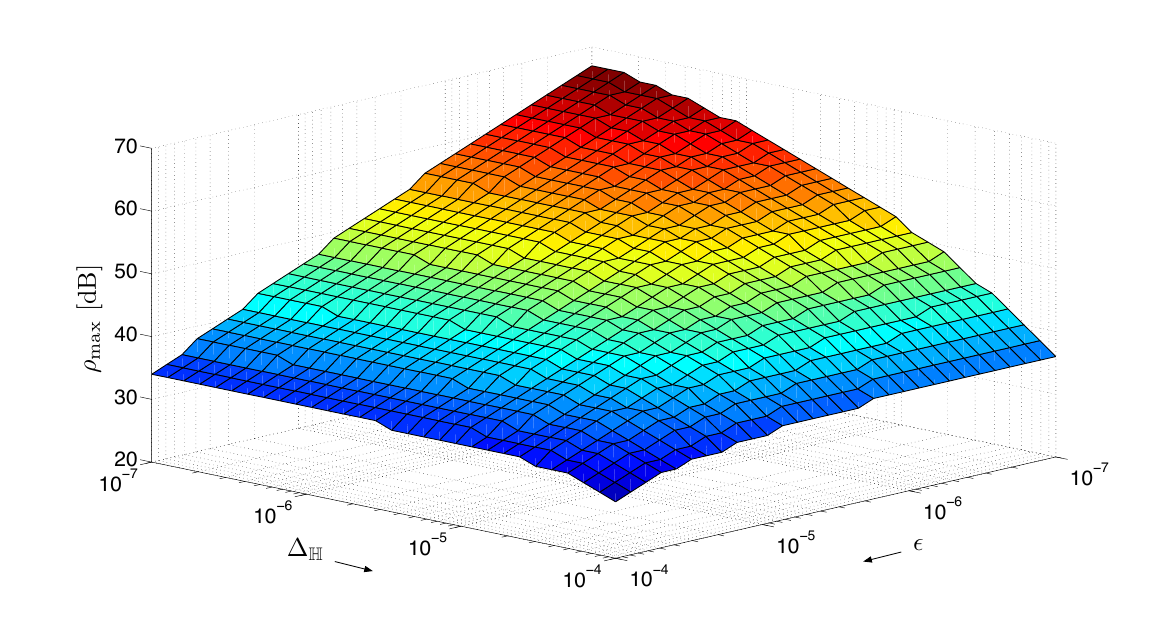}
	\caption{Maximum SNR value~\SNRmax for which~\eqref{eq:accuracy_inequality} holds, as a function of~\spread and~\uncertainty.
	The lower bound~\LBsquare is evaluated for a WH set based on the root-raised-cosine pulse~\eqref{eq:raised_cosine}; furthermore, $\tfstep=1.02$.}
	\label{fig:max}
\end{figure}

The results presented in \fref{fig:lbnormspread4and6}  suggest that, for~$\tfstep=1.02$, the lower bound \LBsquare is close to the AWGN-channel capacity upper bound~\cawgn over a  large range of SNR values. 
To further quantify this statement, we identify the SNR interval~$[\SNRmin,\SNRmax]$ over which
\begin{equation}
\label{eq:accuracy_inequality}
	\LBsquare(\snr)\geq 0.75\,\cawgn(\snr).
\end{equation}
The corresponding interval end points~\SNRmin and~\SNRmax, as a function of~\spread and~\uncertainty,  can easily be obtained numerically and are plotted in 
Figs.~\ref{fig:min} and 
\ref{fig:max}, respectively, for $\tfstep=1.02$.
For the WH set and WSSUS underspread channels considered in this section, we have~$\SNRmin \in [-25\dB, -7\dB]$ and~$\SNRmax \in [30\dB, 68\dB]$. 
Hence, the interval $(\SNRmin,\SNRmax)$ covers all SNR values of practical interest. 
An analytic characterization of~\SNRmin and~\SNRmax seems  difficult. 
Insights on how these two quantities are related to the channel parameters~\spread and~\uncertainty can be obtained by the following ``back-of-the-envelope'' analysis of~$\LBsquare$ (for $\tfstep=1.02$). 
We first approximate \LBsquare by replacing \minambsquare and \maxsumambsquare (whose dependency on \spread is difficult to characterize analytically) with simpler expressions that are accurate when $\spread \ll 1$. 
Then, we determine the SNR values for which the resulting approximate lower bound is close to~\eqref{eq:awgn_cap_approx}.                                                                                 
We start by noting that, when $\spread \ll 1$, we can approximate  \minambsquare by its first-order Taylor-series expansion around $\spread =0$.
This yields
\begin{align}\label{eq:first_order_expansion_minambsquare}
	 \minambsquare&=\min_{(\dd) \in \spreadsquareset} \abs{\afp}^2 \notag\\
	&\approx1-\constm \spread
\end{align}
where $\spreadsquareset\define[-\sqrt{\spread}/2,\sqrt{\spread}/2] \times [-\sqrt{\spread}/2,\sqrt{\spread}/2]$, and~$\constm\define\pi^2(\eftime^2 +\efband^2)$ with
\begin{equation*}
	\eftime^2\define\int\time^2 \abs{\logonp}^2 d\time, \quad
	\efband^2\define\int\freq^2 \abs{\logonfp}^2 d\freq.	
\end{equation*}
To get~\eqref{eq:first_order_expansion_minambsquare}, we used the Taylor-series expansion of  $\abs{\afp}^2$ reported in~\cite[Sec.~6]{wilcox91a}. 
Similarly, for $\spread\ll 1$ we can approximate \maxsumambsquare as follows:    
 \begin{align}
    \label{eq:max_approx}
	\maxsumambsquare&=\max_{(\dd) \in \spreadsquareset} \mathop{\sumdtime\sumdfreq}_{(\dtdf)\neq(0,0)}\abs{\af_{\logon}(\delay -\dtime\srtfstep,\doppler -\dfreq\srtfstep)  }^{2}  \notag \\
	&\approx  \constM \spread
\end{align}
	where 
	\begin{align*}
		 \constM\define\mathop{\sumdtime\sumdfreq}_{(\dtdf)\neq(0,0)}\left[\abs{\dernup}^2+\abs{\dertaup}^2\right]/4
	\end{align*}
	with \dernup and \dertaup being the first partial derivatives of \afp (with respect to \doppler and \delay, respectively) calculated at the points $(-\dtime\srtfstep,-\dfreq\srtfstep)$:
	\begin{align*}
		\dernup&\define-j2\pi \int_{\time} \time \logonp \logon(\time+\dtime\srtfstep)\cex{\dfreq\srtfstep\time}d\time\\
		\dertaup&\define j2\pi \int_{\freq} \freq \logonf(\freq-\dfreq\srtfstep)\logonfp\cexn{\dtime\srtfstep\freq} d\freq.
	\end{align*}  
Here,~\eqref{eq:max_approx} is obtained by performing a Taylor-series expansion of $\af_{\logon}(\delay-\dtime\srtfstep,\doppler-\dfreq\srtfstep)$  around the point $(\delay,\doppler)=(0,0)$ for all \dtime and \dfreq, and by using that $\logonp$ is real and even.
For our choice of $\tfstep=1.02$ we have $\constm\approx25.87$ and $\constM\approx0.77$. 
Hence,~\eqref{eq:first_order_expansion_minambsquare} and~\eqref{eq:max_approx}    suggest that when $\spread\ll 1$, we can approximate \minambsquare by $1$ and \maxsumambsquare by $\spread$. 
On the basis of these two approximations, which are in good agreement with the numerical results reported in \fref{fig:SIR}, and the assumption that $\uncertainty\ll 1$ and $\tfstep =1.02 \approx 1$, we can  approximate the lower bound \LBsquare for all SNR values satisfying $\snr(\spread+\uncertainty)\ll 1$ as follows
\begin{IEEEeqnarray}{rCl}\label{eq:lbsquare_approx}
 	\LBsquare(\SNR)&\approx& \bandwidth\Biggl\{\Ex{\ch}{\log\mathopen{}\left( 1 + \SNR\! \abs{\ch}^{2} \right)}\notag \\
	&&-\:\sqrt{\spread} \log\mathopen{}\left(1 + \frac{\SNR}{\sqrt{\spread}}\right)\Biggr\}.
\end{IEEEeqnarray}
The RHS of~\eqref{eq:lbsquare_approx} is close to the AWGN-channel capacity upper bound (apart from the Jensen penalty in the first term) for all SNR values that satisfy $\SNR \gg \sqrt{\spread}$.
In fact, when $\snr\gg \sqrt{\spread}$ (and $\spread \ll 1$), the second term on the RHS of~\eqref{eq:lbsquare_approx} can be approximated as
\begin{align*}
	\sqrt{\spread} \log\mathopen{}\left(1 + \frac{\SNR}{\sqrt{\spread}}\right) &\approx \sqrt{\spread} \log \snr -\sqrt{\spread} \log \sqrt{\spread}  \\
	&\approx \sqrt{\spread} \log \snr \\
	&\ll \log \snr 
\end{align*}
which implies that, when $\snr\gg \sqrt{\spread}$ (and $\spread \ll 1$), the first term on the RHS of~\eqref{eq:lbsquare_approx} dominates the second term on the RHS of~\eqref{eq:lbsquare_approx}.

We can therefore summarize our findings in the following rule of thumb: the capacity of a Rayleigh-fading WSSUS underspread channel
with scattering function~\scafunp and parameters \spread and \uncertainty  in \fref{def:underspread},
is close to~\cawgn for all \SNR that satisfy $\sqrt{\spread}\ll \SNR \ll 1/(\spread + \uncertainty)$, independently of whether~\scafunp is compactly supported or not, and independently of its shape.
In particular, this implies that capacity essentially grows logarithmically with SNR up to SNR values $\SNR \ll 1/(\spread + \uncertainty)$.
We conclude by noting that the condition $\sqrt{\spread}\ll \SNR \ll 1/(\spread + \uncertainty)$ holds for all channels and SNR values of practical interest.
%
%
%
\section{Conclusions} 
\label{sec:conclusions}
We studied the noncoherent capacity of \emph{continuous-time} Rayleigh-fading channels that satisfy the WSSUS and the underspread assumptions.
Our main result is a capacity lower bound  obtained by (i) discretizing the continuous-time I/O relation and (ii) treating the (signal-dependent) self-interference term in the resulting discretized I/O relation as noise.
Discretization is performed by constraining the input signal to lie in the span of an orthonormal WH set  and by projecting the output signal onto the same orthonormal set.
The resulting lower bound was shown to be close to the AWGN-channel capacity upper bound \cawgn for all SNR values of practical interest, as long as the underlying channel is underspread according to \fref{def:underspread}.
In particular, this result implies that---for all SNR values  typically encountered in real-world systems---the capacity of Rayleigh-fading underspread WSSUS channels is \emph{not sensitive} to whether the channel scattering function is compactly supported or not.   
It also shows that---for all SNR values of practical interest---lack of channel knowledge at the receiver has little impact on the capacity of this class of channels.
From a practical point of view, the underspread assumption is not restrictive as the fading channels commonly encountered in wireless communications are, in fact, highly underspread.

On the basis of our capacity lower bound, we  derived an information-theoretic criterion for the design of capacity-approaching WH sets to be used in PS-OFDM schemes.
This criterion is more fundamental than criteria based on SIR maximization (see \cite{matz07-05a} and references therein), because it sheds light on the trade-off between self-interference reduction and maximization of the number of signal-space dimensions available for communication.
 Unfortunately, the corresponding optimization problem is hard to solve, analytically as well as numerically. 
It turns out, however, that  the simple choice of taking $\logonp$ to be  a root-raised-cosine pulse and letting  the grid-parameter product \tfstep be close to $1$ (but strictly larger than $1$) yields a lower bound that is close to \cawgn for all SNR values of practical interest.                              
In particular, this result suggests that---when self-interference is treated as (signal-dependent) noise---the maximization of the number of signal-space dimensions available for communication should be privileged over SIR maximization.

An interesting open problem, the solution of which would strengthen our results, is to compute an upper bound on the capacity of~\eqref{eq:ltv-io} by assuming perfect channel state information at the receiver.
The main difficulty here lies in dealing with self-interference.
In particular, we expect that nonstandard  
tools from large random matrix theory will be needed for this analysis.   
Recent results along these lines, for a specific channel model, can be found in~\cite{tulino10-03a}.


\appendices

\section{AWGN Capacity Upper Bound} 
\label{app:awgn_capacity_upper_bound}
Let $\CHop \in\chsetp$.                          
To establish that~$\capacity \leq \cawgn$, where \cawgn is defined in~\eqref{eq:awgn_capacity}, we start by upper-bounding the mutual information on the RHS of~\eqref{eq:capacity_definition_ct} as follows:
\begin{equation}\label{eq:chain_rule_for_awgn}	
	\mi(\outpfiltp;\inpp)\leq\mi(\outpfiltp;\outpnnfiltp).
\end{equation}
Here, $\outpnnfiltp\define(\freqtrunc{\bandwidth+2\maxDoppler}\outpnn)(\time)$, and the inequality follows by noting that \inpp and \outpfiltp are conditionally independent given \outpnnfiltp and by using the data-processing inequality for continuous-time random signals~\cite[Thm.~1.4]{gelfand57-a}.
 If we now substitute~\eqref{eq:chain_rule_for_awgn} into~\eqref{eq:capacity_definition_ct}, we obtain
\begin{equation}
	\label{eq:cap_ub_after_chain_rule}
	\capacity\leq 	\lim_{\duration \to \infty} \frac{1}{\duration}\sup_{\probinpset}\mi(\outpfiltp;\outpnnfiltp).
\end{equation}
The mutual information in~\eqref{eq:cap_ub_after_chain_rule}  is between the input and the output of a continuous-time band-limited AWGN channel.  
Hence, we can establish an upper bound on the RHS of~\eqref{eq:cap_ub_after_chain_rule} by invoking~\cite[Thm.~2]{wyner66-03a}, provided that an inequality, in the spirit of~\eqref{eq:spill-over},   on the energy of the restriction of~$\outpnnfilt(\time)$ to a certain time interval can be established. 
More specifically, 
we shall show next that the energy of the restriction of $\outpnnfilt(\time)$ to the interval $[-\duration/2-\maxDelay,\duration/2+\maxDelay]$, i.e., the energy of $(\timetrunc{\duration+2\maxDelay}\outpnnfilt)(\time)$, is bounded from below by $(1-\spillover)(1-\uncertainty)\Ex{}{\vecnorm{\inpp}^2}$.
Let
\begin{equation*}
	\filtinp(\time)\define \freqtrunc{\bandwidth+2\maxDoppler}\mathopen{}\left(\inp(\time-\delay)\cex{\time\doppler}\right).
\end{equation*}
Using
\begin{multline*}	
	(\timetrunc{\duration+2\maxDelay}\outpnnfilt)(\time)\\
	=
	\begin{cases}
		\displaystyle
		\spreadint{\spfp\filtinp(\time)}, & \text{if\,} \abs{\time}\leq \duration/2 +\maxDelay\\
			0, & \text{otherwise}
	\end{cases}
\end{multline*}
we get
\begin{IEEEeqnarray}{rCl}
	\IEEEeqnarraymulticol{3}{l}{\Ex{}{\vecnorm{(\timetrunc{\duration+2\maxDelay}\outpnnfilt)(\time)}^2}} \notag\\
	&\stackrel{(a)}{=}& \spreadint{\scafunp \Ex{}{\int_{-\duration/2-\maxDelay}^{\duration/2+\maxDelay}
	\abs{\filtinp(\time) }^2 d\time}} \notag\\ 
	&\stackrel{(b)}{\geq}& \int_{-\maxDoppler}^{\maxDoppler}\int_{-\maxDelay}^{\maxDelay} 
	\scafunp \Ex{}{\int_{-\duration/2-\maxDelay}^{\duration/2+\maxDelay}
	\abs{\filtinp(\time) }^2 d\time}d\delay d\doppler\IEEEeqnarraynumspace\label{eq:application_WSSUS}
\end{IEEEeqnarray}
where (a) follows from the WSSUS property of $\CHop$ [see~\eqref{eq:scattering_function}], and
(b) follows from the non-negativity of the integrand.
Because $\inpp$ is subject to
 the bandwidth constraint~\eqref{eq:bandwidth_constraint} and to the  time-concentration constraint~\eqref{eq:spill-over}, we have that, for every $(\dd) \in  [-\maxDelay,\maxDelay] \times [-\maxDoppler,\maxDoppler]$,
\begin{equation}
	\label{eq:countering_dispersiveness}	
	\Ex{}{\int_{-\duration/2-\maxDelay}^{\duration/2+\maxDelay}
	\abs{\filtinp(\time) }^2 d\time}
	\geq 
	(1-\spillover)\Ex{}{\vecnorm{\inpp}^2}. 
\end{equation}
Substituting~\eqref{eq:countering_dispersiveness} into~\eqref{eq:application_WSSUS}, we get
\begin{IEEEeqnarray}{rCl}	
	\IEEEeqnarraymulticol{3}{l}{\Ex{}{\vecnorm{(\timetrunc{\duration+2\maxDelay}\outpnnfilt)(\time)}^2}} \notag \\
	\quad&\geq&
	(1-\spillover)\Ex{}{\vecnorm{\inpp}^2}\int_{-\maxDoppler}^{\maxDoppler}\int_{-\maxDelay}^{\maxDelay}\scafunp d\delay d\doppler \notag\\
	\quad&\geq&
		(1-\spillover)(1-\uncertainty)\Ex{}{\vecnorm{\inpp}^2}\label{eq:lb_spillover}
\end{IEEEeqnarray}
where the last step follows from \fref{def:underspread}.
We now observe that 
\begin{align}\label{eq:power_inequality}
   \Ex{}{\vecnorm{\outpnnfiltp}^2} &\leq   \Ex{}{\vecnorm{\outpnnp}^2} \notag \\
&= \spreadint{\scafunp\Ex{}{\vecnorm{\inp(\time-\delay)e^{j2\pi\doppler\time}}^2}} \notag   \\
&=\spreadint{\scafunp\Ex{}{\vecnorm{\inp(\time)}^2}} \notag \\
&=\Ex{}{\vecnorm{\inpp}^2}.
\end{align}
Here, the last step follows from the normalization~\eqref{eq:scafunnorm}.
The inequality~\eqref{eq:power_inequality},  combined with~\eqref{eq:lb_spillover}, yields the following time-concentration inequality  for~\outpnnfiltp   [cf.~\eqref{eq:spill-over}]
\begin{equation}\label{eq:spill_over_awgn}
	\Ex{}{\vecnorm{\timetrunc{\duration+2\maxDelay}\outpnnfiltp}^2}\geq (1-\spillover)(1-\uncertainty)\Ex{}{\vecnorm{\outpnnfiltp}^2}.
\end{equation}

To obtain the desired upper bound~\eqref{eq:awgn_capacity}, we now note that every probability measure on \inpp in the set \probinpset induces a probability measure on \outpnnfiltp (through the map $\outpnnfiltp=(\freqtrunc{\bandwidth+2\maxDoppler}\CHop\inp)(\time)$) that satisfies the following constraints [cf.~\eqref{eq:bandwidth_constraint}--\eqref{eq:av_power_constraint}]:
\begin{enumerate}[i)]
	\item  the bandwidth of \outpnnfiltp is no larger than $(\bandwidth +2 \maxDoppler)$,
	\item $\Ex{}{\vecnorm{\outpnnfiltp}^2}\leq \duration\Pave$, which follows from~\eqref{eq:power_inequality} and~\eqref{eq:av_power_constraint}, and
	\item  \eqref{eq:spill_over_awgn} holds.
\end{enumerate}  
Let~$\altprobinpset$ be the set of \emph{all} probability measures on $\outpnnfiltp$ satisfying i)--iii). 
Note that the set of probability measures on~\outpnnfiltp induced by probability measures on \inpp in \probinpset through the map $\outpnnfiltp=(\freqtrunc{\bandwidth+2\maxDoppler}\CHop\inp)(\time)$ is contained in \altprobinpset, as shown above.
This property can be used to upper-bound the RHS of~\eqref{eq:cap_ub_after_chain_rule} according to
\begin{multline*}
	\lim_{\duration \to \infty} \frac{1}{\duration}\sup_{\probinpset}\mi(\outpfiltp;\outpnnfiltp)\\
	\leq \lim_{\duration \to \infty} \frac{1}{\duration}\sup_{\altprobinpset}\mi(\outpfiltp;\outpnnfiltp).
\end{multline*}
A direct application of~\cite[Thm.~2]{wyner66-03a} yields~\eqref{eq:awgn_capacity}.
\section{The Input Signal~\eqref{eq:canonical-input} Satisfies~\eqref{eq:spill-over}} 
\label{app:proof_of_lemwh_and_spillover}
We show that for every orthonormal WH set satisfying Properties~\ref{prop:bandwidth} and~\ref{prop:decay} in  \fref{sec:a_smaller_set_of_input_distributions} and for every $\spillover>0$, and $\duration>0$, one can find a  $\tslotsguard>0$ such that the corresponding~\inpp in~\eqref{eq:canonical-input} (with $\tslots$ chosen as specified in \fref{sec:time_limitation_constraint_eq:spill-over}) satisfies~\eqref{eq:spill-over}.
To this end, it will turn out convenient to reformulate~\eqref{eq:spill-over} as follows: 
\begin{equation}\label{eq:spill-over-reformulated}	
	\Ex{}{\vecnorm{(\opI-\timetrunc{\duration})\inp(\time)}^2}\leq \spillover \Ex{}{\vecnorm{\inp(\time)}^2}
\end{equation}
where \opI denotes the identity operator. 
Let~\inpvec be the vector of dimension $\tslotstot\fslotstot$ obtained by stacking the data symbols $\inp[\dtdf]$ as in~\eqref{eq:mv_input}.
Furthermore, let
\begin{equation*}	
	\corr[\dtdf,\altdtdf]\define\int_{\abs{\time}>\duration/2} \slogonp\conj{\logon}_{\altdtdf}(\time)  d\time
\end{equation*}
and define $\corrmat$ to be the square matrix of dimension $\tslotstot\fslotstot\times \tslotstot\fslotstot$ with entries
\begin{equation*}	
	[\corrmat]_{\dfreqtilde+\dtimetilde\fslotstot,\altdfreqtilde+\altdtimetilde\fslotstot}\define\corr[\dtimetilde-\tslots,\dfreqtilde-\fslots,\altdtimetilde-\tslots,\altdfreqtilde-\fslots]
\end{equation*}
for $\allz{\dtimetilde,\altdtimetilde}{\tslotstot}$ and  $\allz{\dfreqtilde,\altdfreqtilde}{\fslotstot}$.
Note that~\corrmat is Hermitian, by construction.  
We have that
\begin{equation*}	
	\Ex{}{\vecnorm{(\opI-\timetrunc{\duration})\inp(\time)}^2}=\Ex{}{\herm{\inpvec}\corrmat\inpvec}\leq \eigmax{\corrmat}\Ex{}{\vecnorm{\inpvec}^2}.
\end{equation*}
Here, the first equality follows by definition, and the inequality follows by application of the Rayleigh-Ritz theorem~\cite[Thm.~4.2.2]{horn85a}.\footnote{With  slight abuse of notation, we used $\vecnorm{\cdot}$, a symbol which we reserved for the norm in \hilfunspacep, to  denote  the Euclidean norm in a finite-dimensional vector space.} %
We next use the Ger\v{s}gorin disc theorem~\cite[Cor.~6.1.5]{horn85a} to derive an upper bound
on $\eigmax{\corrmat}$ that is explicit in the entries of \corrmat:
\begin{equation}\label{eq:ub_eigenvalues}
	\eigmax{\corrmat}\leq \max_{\dtime \in [-\tslots,\tslots], \dfreq \in [-\fslots,\fslots]} \left[   \sum_{\altdtime=-\tslots}^{\tslots} \sum_{\altdfreq=-\fslots}^{\fslots} \abs{\corr[\dtdf,\altdtdf]} \right].
\end{equation}
Each term on the RHS of~\eqref{eq:ub_eigenvalues} can be bounded as follows 
\begin{align*}
	\abs{\corr[\dtdf,\altdtdf]} 
	&= \abs{\,\,\int_{\abs{\time}> \duration/2}\slogonp \conj{\logon}_{\altdtdf}(\time)  d\time} \notag \\
	&\leq \int_{\abs{\time}> \duration/2}\abs{\slogonp \conj{\logon}_{\altdtdf}(\time)} d\time \notag \\
	&= \int_{\abs{\time}> \duration/2}\abs{\logon(\time-\dtime\tstep) \conj{\logon}(\time-\altdtime\tstep)} d\time.
\end{align*}
Recall that $\duration/2=(\tslots+\tslotsguard+1/2)\tstep$,  by construction.
As, by assumption,~\logonp is even and satisfies $\logonp=\landauO(1/\time^{1+\decay})$, there exist  constants $\constdecay>0$, and $\time_0>0$ such that $\abs{\logonp}<\constdecay/\abs{\time}^{1+\decay}$ for $\abs{\time}\geq \time_0$.
 Hence, if we choose \tslotsguard such that $\tslotsguard\tstep>\time_0$, we get\footnote{If $\time_0>\duration/2$, we let the guard-interval cover the whole transmission time $[-\duration/2,\duration/2]$.
In this case~\eqref{eq:spill-over-reformulated}  is trivially satisfied.}
\begin{IEEEeqnarray}{rCl}
	 \IEEEeqnarraymulticol{3}{l}
	 {\abs{\corr[\dtdf,\altdtdf]}} \notag \\
	 \quad &\leq& \constdecay^2 \int_{\abs{\time}> (\tslots+\tslotsguard+1/2)\tstep} \frac{1}{\abs{\time-\dtime\tstep}^{1+\decay}}
		\frac{1}{\abs{\time-\altdtime\tstep}^{1+\decay}} d\time\notag\\
			&=&\constdecay^2 \int_{(\tslots+\tslotsguard+1/2)\tstep}^{\infty} \frac{1}{\abs{\time-\dtime\tstep}^{1+\decay}}
			\frac{1}{\abs{\time-\altdtime\tstep}^{1+\decay}} d\time  \notag \\
			&&+\:
			\constdecay^2 \int_{-\infty}^{-(\tslots+\tslotsguard+1/2)\tstep} \frac{1}{\abs{\time-\dtime\tstep}^{1+\decay}}
			\frac{1}{\abs{\time-\altdtime\tstep}^{1+\decay}} d\time
			\notag\\
			&\stackrel{(a)}{\leq}& \constdecay^2 \int_{(\tslots+\tslotsguard+1/2)\tstep}^{\infty} \frac{1}{\abs{\time-\tslots\tstep}^{1+\decay}}
			\frac{1}{\abs{\time-\altdtime\tstep}^{1+\decay}} d\time \notag \\
			&&+\:
			\constdecay^2 \int_{-\infty}^{-(\tslots+\tslotsguard+1/2)\tstep} \frac{1}{\abs{\time+\tslots\tstep}^{1+\decay}}
			\frac{1}{\abs{\time-\altdtime\tstep}^{1+\decay}} d\time \notag\\
			&\stackrel{(b)}{=}& \constdecay^2 \int_{(\tslotsguard+1/2)\tstep}^{\infty} \frac{1}{\abs{\time}^{1+\decay}}
			\frac{1}{\abs{\time-(\altdtime-\tslots)\tstep}^{1+\decay}} d\time \notag \\
			&&+\:
			\constdecay^2 \int_{-\infty}^{-(\tslotsguard+1/2)\tstep} \frac{1}{\abs{\time}^{1+\decay}}
			\frac{1}{\abs{\time-(\altdtime+\tslots)\tstep}^{1+\decay}} d\time.\label{eq:second_bound_app_spillover}
	\end{IEEEeqnarray}
Here, (a) follows by replacing $\dtime$ by $\tslots$ in the first term of the sum and $\dtime$ by $-\tslots$ in the second term of the sum; these substitutions lead to an upper bound; (b) follows by a simple change of variables.
Note now that, for $\time\geq \tslotsguard\tstep$, we have
\begin{align}
	\label{eq:first_sum_finite}
	\sum_{\altdtime=-\tslots}^{\tslots} \frac{1}{\abs{\time-(\altdtime-\tslots)\tstep}^{1+\decay}} 
	&= \sum_{\altdtime=0}^{2\tslots} \frac{1}{\abs{\time+\altdtime\tstep}^{1+\decay}}\notag\\
	&\leq \sum_{\altdtime=0}^{2\tslots} \frac{1}{[(\tslotsguard+\altdtime)\tstep]^{1+\decay}}\notag \\
	&\leq  \sum_{\altdtime=1}^{\infty} \frac{1}{(\altdtime\tstep)^{1+\decay}}\notag \\
	&\define \constdecayalt<\infty
\end{align}
where in the last step we used that~$\decay>0$ and, hence, the series converges.
Similarly, for $\time\leq -\tslotsguard\tstep$, we have
\begin{align}
	\label{eq:second_sum_finite}
\sum_{\altdtime=-\tslots}^{\tslots} \frac{1}{\abs{\time-(\altdtime+\tslots)\tstep}^{1+\decay}} 
&= \sum_{\altdtime=0}^{2\tslots} \frac{1}{\abs{\time-\altdtime\tstep}^{1+\decay}}\notag\\
&\leq \sum_{\altdtime=0}^{2\tslots} \frac{1}{[(\tslotsguard+\altdtime)\tstep]^{1+\decay}}\notag \\
&\leq\constdecayalt.
\end{align}
Inserting~\eqref{eq:second_bound_app_spillover} into~\eqref{eq:ub_eigenvalues} and using~\eqref{eq:first_sum_finite} and~\eqref{eq:second_sum_finite}, we get
\begin{IEEEeqnarray*}{rCl}
	\IEEEeqnarraymulticol{3}{l}
	{\sum_{\altdtime=-\tslots}^{\tslots} \sum_{\altdfreq=-\fslots}^{\fslots}\abs{\corr[\dtdf,\altdtdf]}} \notag\\
	&\leq&
	\sum_{\altdtime=-\tslots}^{\tslots} \sum_{\altdfreq=-\fslots}^{\fslots}\Biggl[\constdecay^2 \int_{(\tslotsguard+1/2)\tstep}^{\infty} \frac{1}{\abs{\time}^{1+\decay}}
	\frac{1}{\abs{\time-(\altdtime-\tslots)\tstep}^{1+\decay}} d\time \\
	&&+
	\constdecay^2 \int_{-\infty}^{-(\tslotsguard+1/2)\tstep} \frac{1}{\abs{\time}^{1+\decay}}
	\frac{1}{\abs{\time-(\altdtime+\tslots)\tstep}^{1+\decay}} d\time\Biggr]\\
	&\leq& 2(2\fslots+1)\constdecay^2\constdecayalt \int_{(\tslotsguard+1/2)\tstep}^{\infty}\frac{1}{\time^{1+\decay}}d\time.
\end{IEEEeqnarray*}
To summarize, we have the following upper bound on the RHS of~\eqref{eq:ub_eigenvalues}:
\begin{equation}
\label{eq:final_bound_app_spillover}
	\eigmax{\corrmat}\leq  2(2\fslots+1)\constdecay^2\constdecayalt \int_{(\tslotsguard+1/2)\tstep}^{\infty}\frac{1}{\time^{1+\decay}}d\time.
\end{equation}
The RHS of~\eqref{eq:final_bound_app_spillover}  can be made arbitrarily small by choosing~\tslotsguard sufficiently large.
In other words,  we can find a finite~\tslotsguard for which the RHS of~\eqref{eq:final_bound_app_spillover} is smaller than~\spillover.
This concludes the proof.

\section{Statistical Properties of the Channel Coefficients in~\eqref{eq:discretized I/O with interference} } 
\label{app:statistical_properties}
We establish basic properties of the statistics of~$\ch[\dtdf]$ and~$\interfp$ in~\eqref{eq:discretized I/O with interference} 
that will be needed in the proof of the capacity lower bound in~\fref{thm:lower bound}.
The first property concerns the autocorrelation function of~$\ch[\dtdf]$. 
Let the \emph{cross-ambiguity function} of two signals~\logonaltp and~\logonp be defined as~\cite{woodward53a}
\begin{align}
\label{eq:ambiguity_function}
 \af_{\logonalt,\logon}(\dd)\define\int_{\time}\logonalt(\time)\conj{\logon}(\time-\delay)
		\cexn{\doppler\time} d\time
\end{align}
and let the ambiguity function of \logonp be defined as~$ \afp\define \af_{\logon,\logon}(\dd)$.\footnote{Basic results on the ambiguity function that will be needed in our analysis are reviewed in~\fref{app:ambiguity}. }
The autocorrelation function of~$\ch[\dtdf]$ turns out to be explicit in the ambiguity function of \logonp, as the following calculation reveals:
\begin{IEEEeqnarray}{rCl}
\IEEEeqnarraymulticol{3}{l}
{\Ex{}{\ch[\dtdf]\conj{\ch}[\altdtdf]}} \notag\\
\quad &=& \Ex{}{\inner{\CHop\slogon}{\slogon}\conj{\inner{\CHop\logon_{\altdtdf}}{\logon_{\altdtdf}}}} \notag\\
\quad&\stackrel{(a)}{=}& \spreadint{\scafunp \conj{\af}_{\slogon}(\dd)\af_{\logon_{\altdtdf}}(\dd)}\notag\\
\quad&\stackrel{(b)}{=}& \spreadint{\scafunp \abs{{\af}_{\logon}(\dd)}^{2} \cex{[(\dtime-\altdtime)\tstep\doppler - (\dfreq-\altdfreq)\fstep\delay]} }\notag\\
\quad&\define&\chcorr[\dtime-\altdtime,\dfreq-\altdfreq].
\label{eq:stationarity}
\end{IEEEeqnarray}
Here,~(a) follows from~\fref{prop:channel operator} in~\fref{app:ambiguity} and because~$\CHop$ is WSSUS [see~\eqref{eq:scattering_function}], while~(b) follows from~\fref{prop:cross-ambiguity} in~\fref{app:ambiguity} [see in particular \eqref{eq:property amb}]. As a consequence of~\eqref{eq:stationarity}, we have that~$\{\ch[\dtdf]\}$ is stationary both in discrete time~\dtime and in discrete frequency~\dfreq.
The corresponding power spectral density function is given by
\begin{align}
\label{eq:chspecfun}
	\chspecfunp\define\sumdtime\sumdfreq
		\chcorr[\Ddtime,\Ddfreq]\cexn{(\Ddtime
		\specparam-\Ddfreq\altspecparam)},\quad \abs{\altspecparam},
		\abs{\specparam}\le 1/2.
\end{align}
The Fourier transform relation~\eqref{eq:chspecfun} together with the Poisson summation formula allow us to relate \chspecfunp to the channel scattering function \scafunp as follows
\begin{IEEEeqnarray}{rCl}
	{\chspecfunp}
	 &=& \sumdtime\sumdfreq \cexn{(\Ddtime\specparam - \Ddfreq
		\altspecparam)} \notag \\
		&&\times \spreadint{\scafunp\abs{{\af}_{\logon}(\dd)}^{2} \cex{(\Ddtime\tstep\doppler -\Ddfreq
		\fstep\delay)}} \notag\\
	&=&\frac{1}{\tfstep}\sumdtime\sumdfreq \scafun\mathopen{}\left(\frac{\altspecparam-\dfreq}{\fstep},\frac{\specparam-\dtime}{\tstep}\right)\notag\\
	&&\times \abs{\af_{\logon}\mathopen{}\left(\frac{\altspecparam-\dfreq}{\fstep},\frac{\specparam-\dtime}{\tstep} \right)}^{2}. 
\label{eq:specfun-scafun}
\end{IEEEeqnarray}
Another property we shall often use is 
\begin{align*}
	\chcorr[0,0]&=\intdiscretetwod \chspecfunp  d\altspecparam d\specparam \notag \\
	&=  \spreadint{\scafunp \abs{{\af}_{\logon}(\dd)}^{2} } \notag \\
	&\leq 1
\end{align*}
where the last step follows from~\fref{prop:ambiguity surface}  in \fref{app:ambiguity}, from the assumption that~\logonp has unit norm, and from the normalization~\eqref{eq:scafunnorm}.

A characterization of the autocorrelation function of~$\interfp$ is possible, but not particularly insightful.  
For our purposes, it will be sufficient to study the variance of $\interfp$. 
As~$\interfp$ has zero mean (see~\fref{sec:the_discretized_i_o_relation}), its variance is given by 
\begin{IEEEeqnarray}{rCl}
\IEEEeqnarraymulticol{3}{l}{\Ex{}{\abs{\interfp}^{2}}} \notag \\
&=& \Ex{}{\abs{\inner{\CHop\logon_{\altdtdf}}{\slogon}}^{2}} \notag\\
&\stackrel{(a)}{=}&\spreadint{ \scafunp \abs{ \af_{\slogon,\logon_{\altdtdf}}(\dd) }^{2} }\notag\\
&\stackrel{(b)}{=}&\spreadint{\scafunp \abs{\af_{g}(\delay + (\altdtime-\dtime)\tstep,\doppler +(\altdfreq-\dfreq)\fstep)  }^{2}} \notag\\
&\define& \intvarp \label{eq:variance interference}
\end{IEEEeqnarray}
where in~(a)  we used~\fref{prop:channel operator} in~\fref{app:ambiguity} together with the WSSUS property of~$\CHop$, and (b) follows from~\fref{prop:cross-ambiguity} in~\fref{app:ambiguity}.
%

%

\section{Properties of the Ambiguity Function} 
\label{app:ambiguity}
We summarize  properties of the (cross-)ambiguity function defined in~\eqref{eq:ambiguity_function} that are needed for our analysis. 
%
%
\begin{prop}\label{prop:ambiguity surface}
		For every function~$\logonp \in \hilfunspacep$,
the ambiguity surface~$\abs{\afp}^2$ attains its maximum  at the origin, i.e.,
	$\abs{\afp}^{2}\le\bigl[\af_{\logon}(0,0)\bigr]^{2}=\vecnorm{\logonp}^4$, for all~\delay and~\doppler.
	This property, as shown in~\cite[Lem. 4.2.1]{groechenig01a}, follows directly from the Cauchy-Schwarz inequality.
\end{prop}
\begin{prop}\label{prop:dilation}
	Let~$\logonp \in \hilfunspacep$ and~$\altaltlogonp=\sqrt{\beta}\logon(\beta\time)$.
	Then
\begin{align*}
		\af_{\altaltlogon}(\dd)&=\int_{\time} \altaltlogonp \conj{\altaltlogon}(\time-\delay)\cexn{\doppler\time} d\time \\
		&=\beta\int_{\time} \logon(\beta\time)\conj{\logon}(\beta(\time-\delay))\cexn{\doppler\time}d\time\\
	&\stackrel{(a)}{=}\int_{z}\logon(z)\conj{\logon}(z-\beta\delay)\cexn{\doppler z/\beta}dz\\
	&=\af_{\logon}\mathopen{}\left(\beta\delay,\frac{\doppler}{\beta}\right)
\end{align*}
	where (a) follows from the change of variables $z=\beta \time$.
\end{prop}

\begin{prop}\label{prop:cross-ambiguity}
The cross-ambiguity function between the two time- and frequency-shifted versions
$\slogonct(\time)\define\logon(\time-\alpha)
\cex{\beta\time}$ and $\slogonctalt(\time)\define\logon(\time-\alpha')
\cex{\beta'\time}$ of~$\logonp \in \hilfunspacep$
is given by
\begin{IEEEeqnarray}{rCl}
	\IEEEeqnarraymulticol{3}{l}{\af_{\slogonct,\slogonctalt}(\dd)} \notag\\
	&=&\int_{\time} \logon(\time-\alpha)\cex{\beta\time}
		\conj{\logon}(\time-\alpha'-\delay)		\cexn{\beta'(\time-\delay)}\cexn{\doppler\time}d\time	\notag\\
	&\stackrel{(a)}{=}&\cex{\beta'\delay} \cexn{(\doppler+\beta'-\beta)\alpha}\notag \\
	&&\times \int_{\time'} \logon(\time')\conj{\logon}(\time'-(\alpha'-\alpha)-\delay)
	\cexn{(\doppler+\beta'-\beta)\time'}	d\time'\notag\\
	&=&\af_{\logon}(\delay+\alpha'-\alpha,\doppler+\beta'-\beta)\notag \\
	&&\times\:\cexn{(\doppler\alpha-\delay\beta')}\cexn{(\beta'-\beta)\alpha}.
	\label{eq:property cross-amb}
\end{IEEEeqnarray}
Here,~(a) follows from the change of variables~$\time'=\time-\alpha$. As a direct consequence of~\eqref{eq:property cross-amb}, we have that
\begin{align}
	\af_{\slogonct}(\dd)=\afp\cexn{(\doppler\alpha-\delay\beta)}.
\label{eq:property amb}
\end{align}
\end{prop}
%
%
\begin{prop}\label{prop:channel operator}
Let~\spfp be the delay-Doppler spreading function of the channel~$\CHop$. Then, for~$\logonp \in \hilfunspacep$, and $\logonaltp \in \hilfunspacep$, we have
\begin{align*}
	\inner{\CHop\logon}{\logonalt}&\stackrel{}{=}\iiint_{\time\,\,\doppler\,\,\delay}\spfp\logon(\time-\delay)\cex{\time\doppler}
	\conj{\logonalt}(\time)d\delay d\doppler d\time\\
	&=\spreadint{\spfp\conj{\Biggl[ \int_{\time}\logonaltp\conj{\logon}(\time-\delay)\cexn{\time\doppler} d\time\Biggr]}}\\
	&=\spreadint{\spfp\conj{\af}_{\logonalt,\logon}(\dd)}.
\end{align*}
\end{prop}

\section{Proof of \fref{thm:lower bound}} 
\label{app:proof_lb}
We obtain a  lower bound on \capacitydisc in~\eqref{eq:capacity_disc} by evaluating the mutual information $\mi(\outpvec;\inpvec)$ for a specific input distribution. 
In particular, we take~$\inp[\dtdf]$ to be \iid JPG with zero mean and variance~$\tfstep\SNR$ for all \dtdf, so that the average-power constraint~\eqref{eq:apc} is satisfied. 
The corresponding input vector~\inpvec is independent of~\chvec,~\interfmat, and~\wgnvec.  
We use the chain rule for mutual information and the fact that mutual information is nonnegative to obtain the following standard lower bound:
\begin{align}
	\mi(\outpvec;\inpvec) &= \mi(\outpvec;\inpvec,\chvec) 
		- \mi(\outpvec;\chvec\given\inpvec)\notag\\
	& = \mi(\outpvec;\chvec)+\mi(\outpvec;\inpvec\given\chvec) -
		\mi(\outpvec;\chvec\given\inpvec)\notag\\
	&\ge \mi(\outpvec;\inpvec\given\chvec) -
		\mi(\outpvec;\chvec\given\inpvec).
	\label{eq:lb-mi-decomp}	
\end{align}
The first term on the RHS of~\eqref{eq:lb-mi-decomp} can be interpreted as a ``coherent'' mutual information term (i.e., the mutual information between \inpvec and \outpvec under perfect knowledge of the channel realization at the receiver), while the second term can be interpreted as quantifying the rate penalty due to the lack of channel knowledge~\cite{durisi10-01a}.

\subsubsection{The ``Coherent'' Term} 
\label{sec:the_coherent_term}
The first term can be further lower-bounded as follows
\begin{IEEEeqnarray*}{rCl}
	\IEEEeqnarraymulticol{3}{l}{\mi(\outpvec;\inpvec\given\chvec)} \\
	\quad&=& \difentp{\inpvec \given \chvec} - \difentp{\inpvec \given \chvec, \outpvec} \\
	\quad&\stackrel{(a)}{=}& \difentp{\inpvec}- \difentp{\inpvec \given \chvec, \outpvec}\\
	\quad&\stackrel{(b)}{=}& \sumdtimelimited\sumdfreqlimited\Bigl[ \difentp{\inp[\dtdf]\given  \inpvecpred } \notag \\
	\quad &&-\difentp{\inp[\dtdf]\given  \chvec, \outpvec,  \inpvecpred } \Bigr]\\
	\quad&\stackrel{(c)}{=}&\sumdtimelimited\sumdfreqlimited\left[ \difentp{\inp[\dtdf] } - \difentp{\inp[\dtdf]\given  \chvec, \outpvec,  \inpvecpred}\right] \\
	\quad&\stackrel{(d)} {\geq}&\sumdtimelimited\sumdfreqlimited\left[ \difentp{\inp[\dtdf] } - \difentp{\inp[\dtdf]\given  \ch[\dtdf], \outp[\dtdf]}\right] \\
	\quad&=&\sumdtimelimited\sumdfreqlimited\mi(\outp[\dtdf];\inp[\dtdf]\given \ch[\dtdf]).
\end{IEEEeqnarray*}
Here,~(a) follows because~\inpvec and~\chvec are independent; (b) is a consequence of the chain rule for differential entropy [\inpvecpred denotes the  vector containing all entries of~$\inpvec$ up to and including the one before $\inp[\dtime,\dfreq]$]. 
Next,~(c) holds because~\inpvec has \iid entries, and~(d) follows because conditioning reduces entropy.

We next seek a lower bound on~$\mi(\outp[\dtdf];\inp[\dtdf]\given \ch[\dtdf])$ that does not depend on~$[\dtdf]$. 
Let~$\altwgnp$ be the sum of the self-interference and noise terms in~$\outp[\dtdf]$ [see~\eqref{eq:discretized I/O with interference}], i.e.,
\begin{equation*}
	\altwgnp \define\mathop{\altsumdtimelimited\altsumdfreqlimited}_{(\altdtdf)\neq(\dtdf)} \interfp \inp[\altdtdf]+\wgn[\dtdf].
\end{equation*}
Furthermore, let~$\altwgnpgauss$ be a proper Gaussian random variable that has the same variance as~\altwgnp. 
It follows from~\cite[Lem.~II.2]{diggavi01-11a} that~$\mi(\outp[\dtdf];\inp[\dtdf]\given \ch[\dtdf])$ does not increase if we replace~$\wgn[\dtdf]$ by~$\altwgnpgauss$. 
Hence, 
\begin{IEEEeqnarray}{rCl}
	\IEEEeqnarraymulticol{3}{l}
	{\mi(\outp[\dtdf];\inp[\dtdf]\given \ch[\dtdf])} \notag \\
	\quad&=&\mi(\ch[\dtdf]\inp[\dtdf]+\altwgnp;\inp[\dtdf]\given \ch[\dtdf]) \notag\\
	\quad&\geq& \mi(\ch[\dtdf]\inp[\dtdf]+\altwgnpgauss;\inp[\dtdf]\given \ch[\dtdf]) \notag\\
	\quad&\stackrel{(a)}{=}& \Ex{\ch[\dtdf]}{	\log\mathopen{}\left( 1 +\tfstep\SNR \frac{\abs{\ch[\dtdf]}^{2}}{\Ex{}{\abs{\altwgnpgauss}^{2}}} \right)	} \notag\\
	\quad&\stackrel{(b)}{=}& \Ex{\ch}{	\log\mathopen{}\left( 1 + \frac{\chcorr[0,0]\tfstep\SNR\abs{\ch}^{2}}{\Ex{}{\abs{\altwgnpgauss}^{2}}} \right)	}
	\label{eq:diggavi bound}
\end{IEEEeqnarray}
where~(a) follows because~$\inp[\dtdf]\distas\jpg(0,\tfstep\SNR)$,
and~(b) follows because~$\ch[\dtdf]\distas\jpg(0,\chcorr[0,0])$ [see~\eqref{eq:stationarity}], so that we can replace~$\ch[\dtdf]$ by~$\chcorr[0,0]\ch$, where~$\ch\distas \jpg(0,1)$. 
As the input symbols~$\inp[\dtdf]$ are independent, and as~$\Ex{}{\abs{\interfp}^2}=\intvarp$ [see~\eqref{eq:variance interference}], we have that
\begin{align}
\label{eq:variance equivalent noise}
	\Ex{}{\abs{\altwgnpgauss}^{2}}&=\Ex{}{\abs{\altwgnp}^{2}} \notag \\
	&= 1 + \tfstep\SNR \mathop{\altsumdtimelimited\altsumdfreqlimited}_{(\altdtdf)\neq(\dtdf)}  \intvarp.
\end{align}
The nonnegativity of~$\intvar[\dtdf]$ allows us to upper-bound~\eqref{eq:variance equivalent noise} as follows
\begin{align}
\label{eq: variance eq. noise ub}
\Ex{}{\abs{\altwgnpgauss}^{2}}&\leq 1 + \tfstep\SNR \mathop{\sum_{\altdtime=-\infty}^{\infty}\sum_{\altdfreq=-\infty}^{\infty}}_{(\altdtdf)\neq(\dtdf)}  \intvarp \notag\\
&= 1 + \tfstep\SNR \mathop{\sum_{\altdtime=-\infty}^{\infty}\sum_{\altdfreq=-\infty}^{\infty}}_{(\altdtdf)\neq(0,0)}  \intvar[\altdtdf] \notag \\
&= 1 + \tfstep\SNR\,\intvartot
\end{align}
where we set
\begin{equation}
\label{eq:def of intvartot}
\displaystyle \intvartot \define \mathop{\sum_{\altdtime=-\infty}^{\infty}\sum_{\altdfreq=-\infty}^{\infty}}_{(\altdtdf)\neq(0,0)}  \intvar[\altdtdf].
\end{equation}
If we now substitute~\eqref{eq: variance eq. noise ub} into~\eqref{eq:diggavi bound}, we obtain
\begin{equation*}
 	\mi(\outp[\dtdf];\inp[\dtdf]\given \ch[\dtdf])\ge  \Ex{\ch}{	\log\mathopen{}\left( 1 + \frac{\chcorr[0,0]\tfstep\SNR\abs{\ch}^{2}}{1 + \tfstep\SNR\,\intvartot } \right)	} 
\end{equation*}
and, consequently,
\begin{equation}
\label{eq:final bound first term}
	\mi(\outpvec;\inpvec\given\chvec) \ge \tslotstot\fslotstot \Ex{\ch}{	\log\mathopen{}\left( 1 + \frac{\chcorr[0,0]\tfstep\SNR\abs{\ch}^{2}}{1 + \tfstep\SNR\,\intvartot } \right)	} .
\end{equation}

\subsubsection{The Penalty Term} 
\label{sec:the_penalty_term}
We next seek an upper bound on the penalty term~$\mi(\outpvec;\chvec\given\inpvec)$ in~\eqref{eq:lb-mi-decomp}. 
The main difficulty lies in the self-interference term being signal-dependent.
Our approach is to split  \outpvec into a self-interference-free part and a self-interference-only part.
Specifically, let~$\wgnvecone\distas\jpg(\veczero,\noisesplit\matI)$ and~$\wgnvectwo\distas\jpg(\veczero,(1-\noisesplit)\matI)$, where $0<\noisesplit<1$, be two~$\tslotstot\fslotstot$-dimensional independent JPG vectors.\footnote{The role of \noisesplit will become clear later.} 
Then,
\begin{align*}
 	\outpvec&=\chvec\had\inpvec+ \interfmat\inpvec +\wgnvec \\
	&= \underbrace{\chvec\had \inpvec+\wgnvecone}_{\define\, \outpvecone} +\underbrace{\interfmat\inpvec +\wgnvectwo}_{\define\, \outpvectwo}.
\end{align*}
By the data-processing inequality~\cite[Thm. 2.8.1]{cover06-a} and the chain rule for mutual information, we have that
\begin{align}
\label{eq:split of the penalty term}
	\mi(\outpvec;\chvec\given \inpvec) &\leq \mi(\outpvecone,\outpvectwo; \chvec \given \inpvec) \notag\\
	&= \mi(\outpvecone;\chvec\given \inpvec) 
	+ \mi(\outpvectwo;\chvec \given \inpvec,\outpvecone).
\end{align}
As~\chvec is JPG, the first term on the RHS of~\eqref{eq:split of the penalty term} can be bounded as follows:
\begin{IEEEeqnarray}{rCl}
	\IEEEeqnarraymulticol{3}{l}{\mi(\outpvecone;\chvec\given \inpvec)} \notag\\
	\quad&=& \mi(\chvec\had\inpvec+\wgnvecone;\chvec \given \inpvec)\notag\\
	\quad&=&\Ex{\inpvec}{\logdet{\matI + \frac{1}{\noisesplit} \diag\{\inpvec\}\Ex{}{\chvec\herm{\chvec}}\diag\{\herm{\inpvec}\}	 }} \notag\\
	\quad&\stackrel{(a)}{=}& \Ex{\inpvec}{\logdet{\matI + \frac{1}{\noisesplit} \diag\{\herm{\inpvec}\}\diag\{\inpvec\}\Ex{}{\chvec\herm{\chvec}}	 }} \notag \\ 
	\quad&\stackrel{(b)}{\le}& \logdet{ \matI + \frac{\tfstep\SNR}{\noisesplit} \Ex{}{\chvec\herm{\chvec}} }.\label{eq:ub penalty 1}
\end{IEEEeqnarray}
Here,~(a) follows from the identity~$\det\mathopen{}\left(\imat + \matA
\herm{\matB}\right) = \det\mathopen{}\left(\imat +\herm{\matB}\matA\right)$ 
for any pair of matrices~$\matA$ and~$\matB$ of appropriate
dimensions~\cite[Thm.~1.3.20]{horn85a} and~(b) is a consequence of Jensen's inequality.

For the second term on the RHS of~\eqref{eq:split of the penalty term} we note that
\begin{align*}
	\mi(\outpvectwo;\chvec \given \inpvec,\outpvecone)&=\difent(\outpvectwo \given \inpvec,\outpvecone) - \difent(\outpvectwo \given \inpvec,\outpvecone,\chvec) \\
	&\stackrel{(a)}{=}\difent(\outpvectwo \given \inpvec,\outpvecone) - \difent(\outpvectwo \given \inpvec,\chvec) \\
    &\stackrel{(b)}{\le} \difent(\outpvectwo \given \inpvec) - \difent(\outpvectwo \given \inpvec,\chvec,\interfmat) \\
    &\stackrel{(c)}{=} \difent(\outpvectwo \given \inpvec) - \difent(\outpvectwo \given \inpvec,\interfmat) \\
    &=\mi(\outpvectwo;\interfmat\given \inpvec).
\end{align*}
Here,~(a) holds because~\outpvecone and~\outpvectwo are conditionally independent given~\inpvec and~\chvec, in~(b) we used twice that conditioning reduces entropy, and~(c) follows because~\outpvectwo and~\chvec are conditionally independent given~\interfmat.

Let~$\corrintinp\define\Ex{\interfmat}{\interfmat\inpvec\herm{\inpvec}\herm{\interfmat}}$ be the~$\tslotstot\fslotstot \times \tslotstot\fslotstot$ conditional covariance matrix of the vector~$\interfmat\inpvec$ given \inpvec. We next upper-bound~$\mi(\outpvectwo;\interfmat\given \inpvec)$ as follows:
\begin{IEEEeqnarray*}{rCl}
	\IEEEeqnarraymulticol{3}{l}{\mi(\outpvectwo;\interfmat\given \inpvec)} \\
	&=&\mi(\interfmat\inpvec+\wgnvectwo; \interfmat\given \inpvec)\\
	&\stackrel{(a)}{=}&
	\Ex{\inpvec}{\logdet{\matI + \frac{1}{1-\noisesplit} \corrintinp} } \\
	&\stackrel{(b)}{\le}&\sum_{\dtimetilde=0}^{\tslotstot-1}\sum_{\dfreqtilde=0}^{\fslotstot-1}\Ex{\inpvec}{\log\mathopen{}\left(1 + \frac{1}{1-\noisesplit} \left[\corrintinp \right]_{(\dfreqtilde+\dtimetilde\fslotstot,\dfreqtilde+\dtimetilde\fslotstot)}		\right)} \\
	&\stackrel{(c)}{\le}& \sum_{\dtimetilde=0}^{\tslotstot-1}\sum_{\dfreqtilde=0}^{\fslotstot-1}\log\mathopen{}\left(1 + \frac{1}{1-\noisesplit} \Ex{\inpvec}{\left[\corrintinp \right]_{(\dfreqtilde+\dtimetilde\fslotstot,\dfreqtilde+\dtimetilde\fslotstot)}}\right)
\end{IEEEeqnarray*}
where~(a) follows because, given \inpvec, the vector~$\interfmat\inpvec$ is JPG, in~(b) we used Hadamard's inequality, and~(c) follows from Jensen's inequality.
As the entries of~\inpvec are~\iid with zero mean, we have that
\begin{IEEEeqnarray*}{rCl}
	\IEEEeqnarraymulticol{3}{l}{\Ex{\inpvec}{\left[\corrintinp\right]_{(\dfreqtilde+\dtimetilde\fslotstot,\dfreqtilde+\dtimetilde\fslotstot)}}} \\
	\quad &=&
	\Ex{\inpvec}{\Ex{\interfmat}{\left[\interfmat\inpvec\herm{\inpvec}\herm{\interfmat}\right]_{(\dfreqtilde+\dtimetilde\fslotstot,\dfreqtilde+\dtimetilde\fslotstot)}}}\\
	\quad &=& \tfstep\SNR \mathop{\altsumdtimelimited\altsumdfreqlimited}_{(\altdtdf)\neq(\dtimetilde-\tslots,\dfreqtilde-\fslots)} 
\intvar[\altdtime-\dtimetilde+\tslots,\altdfreq-\dfreqtilde+\fslots] \\
  \quad	&\le& \tfstep\SNR\, \intvartot
\end{IEEEeqnarray*}
where~\intvartot was defined in~\eqref{eq:def of intvartot}.
Hence,
\begin{equation}
\label{eq:ub penalty 2}
	\mi(\outpvectwo;\interfmat\given \inpvec) \leq \tslotstot\fslotstot \log\mathopen{}\left(1 + \frac{\tfstep\SNR}{1-\noisesplit}\intvartot\right).
\end{equation}
If we now substitute~\eqref{eq:ub penalty 1} and~\eqref{eq:ub penalty 2} into~\eqref{eq:split of the penalty term}, we obtain
\begin{IEEEeqnarray}{rCl}
\label{eq:final bound penalty}  
	\mi(\outpvec;\chvec\given \inpvec) &\leq&  \logdet{ \matI + \frac{\tfstep\SNR}{\noisesplit} \Ex{}{\chvec\herm{\chvec}} }\notag\\
	&&+\tslotstot\fslotstot \log\mathopen{}\left(1 + \frac{\tfstep\SNR}{1-\noisesplit}\intvartot\right).
\end{IEEEeqnarray}
%

\subsubsection{Putting the Pieces Together} 
\label{sec:a_first_lower_bound}

We substitute~\eqref{eq:final bound first term} and~\eqref{eq:final bound penalty} into~\eqref{eq:lb-mi-decomp} and then~\eqref{eq:lb-mi-decomp} into~\eqref{eq:capacity_disc}  to get the following lower bound on capacity:
\begin{multline*}
\capacity(\SNR) 
\geq \frac{\fslotstot}{\tstep} 
\Ex{\ch}{	\log\mathopen{}\left( 1 +\frac{ \chcorr[0,0]\tfstep\SNR\abs{\ch}^{2}}{1 + \tfstep\SNR\,\intvartot } \right)} \\
- \Biggl\{\limintime \frac{1}{(\tslotstot+2\tslotsguard)\tstep}\logdet{ \matI + \frac{\tfstep\SNR}{\noisesplit} \Ex{}{\chvec\herm{\chvec}} }\\
+\frac{\fslotstot}{\tstep} \log\mathopen{}\left(1 + \frac{\tfstep\SNR}{1-\noisesplit}\intvartot\right) \Biggr\}.
\end{multline*}
Furthermore, as the bound holds for all $\noisesplit \in (0,1)$, we can tighten it according to
\begin{multline}
\label{eq:firstLB}
\capacity(\SNR) 
\geq \frac{\fslotstot}{\tstep} 
\Ex{\ch}{	\log\mathopen{}\left( 1 +\frac{ \chcorr[0,0]\tfstep\SNR\abs{\ch}^{2}}{1 + \tfstep\SNR\,\intvartot } \right)} \\
- \inf_{0<\noisesplit<1}\Biggl\{ \limintime \frac{1}{(\tslotstot+2\tslotsguard)\tstep}\logdet{ \matI + \frac{\tfstep\SNR}{\noisesplit} \Ex{}{\chvec\herm{\chvec}} }\\
+\frac{\fslotstot}{\tstep} \log\mathopen{}\left(1 + \frac{\tfstep\SNR}{1-\noisesplit}\intvartot\right) \Biggr\}.
\end{multline}
By direct application
of~\cite[Thm.~3.4]{miranda00-02a}, an extension of Szeg\"o's theorem 
(on the asymptotic eigenvalue distribution of Toeplitz matrices) to two-level Toeplitz matrices, we obtain
\begin{multline*}
	\limintime \frac{1}{(\tslotstot+2\tslotsguard)\tstep}\logdet{ \matI + \frac{\tfstep\SNR}{\noisesplit} \Ex{}{\chvec\herm{\chvec}} } \\
	=\frac{1}{\tstep}\intdiscrete \logdet{\matI + \frac{\tfstep\SNR}{\noisesplit}\chspecfunmatp} d\specparam. 
\end{multline*}
Substituting this expression into~\eqref{eq:firstLB} and noting that [see~\eqref{eq:variance interference}]
\begin{align}
	\intvartot&=
	\mathop{\sum_{\dtime=-\infty}^{\infty}\sum_{\dfreq=-\infty}^{\infty}}_{(\dtdf)\neq(0,0)}  \intvar[\dtdf] \notag\\
	&=\mathop{\sum_{\dtime=-\infty}^{\infty}\sum_{\dfreq=-\infty}^{\infty}}_{(\dtdf)\neq(0,0)} \spreadint{\scafunp \abs{\af_{\logon}(\delay -\dtime\tstep,\doppler -\dfreq\fstep)  }^{2}}
\end{align}
completes the proof.

\section{Proof of~\fref{cor:LB}} 
\label{app:a_lower_bound_explicit_in_spread_and_uncertainty}
To prove the corollary we further  bound each term in~\eqref{eq:lower bound explicit in scattering function} separately.  

\paragraph{The~``$\log\det$'' term} 

We start with an upper bound on the~``$\log\det$'' term on the RHS of~\eqref{eq:lower bound explicit in scattering function}. 
The matrix~\chspecfunmatp is Toeplitz [see~\eqref{eq:mvspec}]. 
Hence, the entries on the main diagonal of~\chspecfunmatp are all equal. Let~\chspecfunentrypzero denote one such entry; then
\begin{align}
\chspecfunentrypzero
&\stackrel{(a)}{=}\sumdtime \chcorr[\dtime,0]\cexn{\dtime\specparam}\notag\\
&\stackrel{(b)}{=}\intdiscrete\chspecfunp d\altspecparam, \,\quad\abs{\specparam}\leq 1/2.
\label{eq:diagonal term of specral matrix}
\end{align}
Here,~(a) follows from~\eqref{eq:mvspec} and~\eqref{eq:stationarity};~(b) follows from~\eqref{eq:chspecfun} and by applying the Poisson summation formula. 
By Hadamard's inequality, we can upper-bound the ``$\log\det$'' term on the RHS of~\eqref{eq:lower bound explicit in scattering function} as follows:
\begin{multline}
\label{eq:Hadamard step}
\frac{1}{\tstep}\intdiscrete \logdet{\matI + \frac{\tfstep\SNR}{\noisesplit}\chspecfunmatp} d\specparam \\ 
\leq \frac{\fslotstot}{\tstep}   \intdiscrete \log\mathopen{}\left( 1 +  \frac{\tfstep\SNR}{\noisesplit}  \chspecfunentrypzero\right) d\specparam.
\end{multline}
Let $\setB\define\{\theta \sothat \abs{\theta}< \maxDoppler\tstep\}$ and $\bar{\setB}\define \{\theta \sothat \maxDoppler\tstep<\abs{\theta}<1/2 \}$.
We next use that~$\maxDoppler\tstep<1/2$, by assumption, to first split the integral into two parts and then use Jensen's inequality on both terms to obtain
\begin{IEEEeqnarray}{rCl}
	\IEEEeqnarraymulticol{3}{l}{\frac{\fslotstot}{\tstep}\intdiscrete \log\mathopen{}\left( 1 +  \frac{\tfstep\SNR}{\noisesplit}  \chspecfunentrypzero\right) d\specparam} \notag \\ 
	&=&  \frac{\fslotstot}{\tstep}\intspecparamin{
	\log\mathopen{}\left( 1 +  \frac{\tfstep\SNR}{\noisesplit}  \chspecfunentrypzero\right)}\notag\\ 
	&&+\:\frac{\fslotstot}{\tstep} \intspecparamout{ \log\mathopen{}\left( 1 +  \frac{\tfstep\SNR}{\noisesplit}  \chspecfunentrypzero\right)} \notag\\
	 &\leq& 2\maxDoppler\fslotstot\log\mathopen{}\left( 1 +\frac{\fstep\SNR}{2\maxDoppler\noisesplit}  \intspecparamin{  \chspecfunentrypzero}  \right) 
	 + \frac{\fslotstot}{\tstep}(1-2\maxDoppler\tstep) \notag \\
	 &&\times 
	 \log\mathopen{}\left( 1 +\frac{\tfstep\SNR}{(1-2\maxDoppler\tstep)\noisesplit} \intspecparamout{  \chspecfunentrypzero}\right).
	\label{eq:twoIntegrals}
\end{IEEEeqnarray}
Let $F(\delay,\doppler)\define \scafunp\abs{\afp}^{2}$.
Note that
\begin{IEEEeqnarray}{rCl}
\IEEEeqnarraymulticol{3}{l}
{\intspecparamin{\!\!\chspecfunentrypzero}} \notag \\
 &\stackrel{(a)}{=}&\intdiscrete\intspecparamin{\!\!\chspecfunp} d\altspecparam \notag\\
&\stackrel{(b)}{=}&\intdiscrete\intspecparamin{\frac{1}{\tfstep}\sumdtime
	\sumdfreq F\mathopen{}\left(\frac{\altspecparam - \Ddfreq }{\fstep},\frac{\specparam -
	\Ddtime}{\tstep}\right) 
}d\altspecparam \notag\\
&\stackrel{(c)}{\leq}& \frac{1}{\tfstep}\intdiscrete\intspecparamin{\sumdtime
	\sumdfreq \scafun\mathopen{}\left(\frac{\altspecparam - \Ddfreq }{\fstep},\frac{\specparam -
	\Ddtime}{\tstep}\right)}d\altspecparam\notag\\
	&\leq&	\frac{1}{\tfstep}\intdiscrete\intdiscrete\sumdtime
		\sumdfreq \scafun\mathopen{}\left(\frac{\altspecparam - \Ddfreq }{\fstep},\frac{\specparam -
		\Ddtime}{\tstep}\right)d \specparam d\altspecparam \notag\\
		&=&\spreadint{\scafunp}=1
		\label{eq:specparamin_integral}
\end{IEEEeqnarray}
where~(a) follows from~\eqref{eq:diagonal term of specral matrix},~(b) follows from~\eqref{eq:specfun-scafun}, and~(c) follows from~\fref{prop:ambiguity surface} in~\fref{app:ambiguity}.
Similar steps lead to
\begin{IEEEeqnarray}{rCl}
\IEEEeqnarraymulticol{3}{l}{\intspecparamout{\chspecfunentrypzero}}\notag \\
&\leq& \frac{1}{\tfstep}\intdiscrete\intspecparamout{\sumdtime
	\sumdfreq \scafun\mathopen{}\left(\frac{\altspecparam - \Ddfreq }{\fstep},\frac{\specparam -
	\Ddtime}{\tstep}\right)}d\altspecparam \notag \\
	&\leq& \int_{\abs{\doppler}\geq \maxDoppler}\int_{\delay} \scafunp  d\delay d\doppler \leq  \uncertainty
\label{eq:specparamout_integral}
\end{IEEEeqnarray}
where the last step follows from~\eqref{eq:underspread_definition}.
If we now substitute~\eqref{eq:specparamin_integral} and~\eqref{eq:specparamout_integral} into~\eqref{eq:twoIntegrals}, insert the result into~\eqref{eq:Hadamard step}, set~$\altspread=2\maxDoppler\tstep$, and use~$\bandwidth=\fslotstot\fstep$, we get
\begin{IEEEeqnarray}{rCl}
 \IEEEeqnarraymulticol{3}{l}{\frac{1}{\tstep}\intdiscrete \logdet{\matI + \frac{\tfstep\SNR}{\noisesplit}\chspecfunmatp} d\specparam}  \notag\\ 
 \qquad&\leq&\frac{\bandwidth\altspread}{\tfstep}\log\mathopen{}\left(1 + \frac{\tfstep\SNR}{\noisesplit\altspread} \right) \notag\\
 \qquad &&+\: \frac{\bandwidth}{\tfstep}(1-\altspread)\log\mathopen{}\left(1 +\frac{\tfstep\SNR\uncertainty}{\noisesplit(1-\altspread)} \right).
 \label{eq:upper_bound_on_logdet}
\end{IEEEeqnarray}

\paragraph{Bounds on~$\chcorr[0,0]$ and on~\intvartot} 
\label{sec:bounds_on_chcorr_0_0}

To further lower-bound the RHS of~\eqref{eq:lower bound explicit in scattering function}, we next derive a lower bound on~$\chcorr[0,0]$ and an upper bound on~\intvartot;
the resulting bounds are explicit in~\spread and~\uncertainty, and in the ambiguity function of~\logonp. 

Let~$\spreadset=\{ (\delay,\doppler) \in [-\maxDelay,\maxDelay]\times [-\maxDoppler,\maxDoppler]\}$ be the rectangular area in the delay-Doppler plane that supports at least~$1-\uncertainty$ of the volume of~\scafunp according to~\eqref{eq:underspread_definition}. 
The following chain of inequalities holds:
\begin{align}
	\chcorr[0,0]&=\spreadint{\scafunp\abs{\afp}^2} \notag\\
	&\geq \iint_{\spreadset} \scafunp\abs{\afp}^2 d\delay d \doppler\notag\\
	&\geq \min_{(\dd) \in \spreadset} \Bigl\{\abs{\afp}^2\Bigr\} \iint_{\spreadset}\scafunp d\delay d\doppler\notag \\
	&\geq  \min_{(\dd) \in \spreadset} \Bigl\{\abs{\afp}^2\Bigr\} (1-\uncertainty).
	\label{eq: lb_on_chcorr}
\end{align}

We now seek an upper bound on~\intvartot. 
Let 
\begin{equation*}
\sumambsquarep=\mathop{\sumdtime\sumdfreq}_{(\dtdf)\neq(0,0)}\abs{\af_{\logon}(\delay -\dtime\tstep,\doppler -\dfreq\fstep)  }^{2}	
\end{equation*}
and note that
\begin{align}
\sumambsquarep&\leq  \sumdtime\sumdfreq \abs{\af_{\logon}(\delay -\dtime\tstep,\doppler -\dfreq\fstep)  }^{2} \notag\\
&= \sumdtime\sumdfreq \abs{\inner{\logon(\time+\delay)\cexn{\doppler\time}}{\slogonp}}^2 \notag\\
&\stackrel{(a)}{\leq} \vecnorm{\logonp}^2=1
\label{eq:max_sumambsquare}
\end{align}
where~(a) follows from Bessel's inequality~\cite[Thm. 3.4-6]{kreyszig89}. The following chain of inequalities holds:
\begin{IEEEeqnarray}{rCl}
\label{eq:ub_on_totinf}
	\intvartot&=&\spreadint{\scafunp\sumambsquarep} \notag\\
	&=&\iint_{\spreadset}\scafunp\sumambsquarep d\delay d\doppler +\iint_{\reals^2\setminus \spreadset}\scafunp \sumambsquarep d\delay d\doppler\notag\\
	&\leq& \max_{(\dd) \in \spreadset} \Bigl\{\sumambsquarep\Bigr\} \iint_{\spreadset}\scafunp d\delay d\doppler \notag \\
	&&+ \max_{(\delay,\doppler) \in \reals^2 \setminus \spreadset}\Bigl\{\sumambsquarep\Bigr\} \iint_{\reals^2\setminus\spreadset}\scafunp d\delay d\doppler\notag\\
	&\stackrel{(a)}{\leq}&   \max_{(\dd) \in \spreadset} \Bigl\{\sumambsquarep\Bigr\} + \uncertainty
\end{IEEEeqnarray}
where~(a) follows from~\eqref{eq:max_sumambsquare},~\eqref{eq:scafunnorm}, and~\eqref{eq:underspread_definition}.
%

The proof is completed by substituting ~\eqref{eq:upper_bound_on_logdet}
into~\eqref{eq:lower bound explicit in scattering function}, and using~\eqref{eq: lb_on_chcorr} and~\eqref{eq:ub_on_totinf} in~\eqref{eq:lower bound explicit in scattering function}.



\section{Proof of \fref{lem:square_setting}} 
\label{app:proof_of_lem:square_setting}
	To prove the lemma, we verify that after the substitutions
	\begin{align*}
		\altaltlogonp&=\sqrt{\beta}\logon(\beta \time)\\
		\alttstep&=\tstep/\beta\\
		\altfstep&=\beta\fstep\\
		\altmaxDelay&=\maxDelay/\beta\\
		\altmaxDoppler&=\beta\maxDoppler
	\end{align*}
	the lower bound~\LBsimple in~\eqref{eq:capacity_lower_bound_explicit_in_spread_and_uncertainty}
	does not change. 
	Note first that
	$\alttstep\altfstep=\tfstep$ and~$\maxDoppler\tstep=\altmaxDoppler\alttstep$. 
Furthermore,~$\vecnorm{\altaltlogonp}=\vecnorm{\logonp}=1$ and, by~\fref{prop:dilation} in~\fref{app:ambiguity}, the orthonormality of~$\{\slogonp\}$ implies the orthonormality of~$\{\altaltlogon(\time-\dtime\alttstep)\cex{\dfreq\altfstep\time}\}$. 
Let now~$\altspreadset=[-\altmaxDelay,\altmaxDelay]\times[-\altmaxDoppler,\altmaxDoppler]$; we have that 
\begin{align*}
	\minambsquare&=\min_{(\dd) \in \spreadset}\abs{\afp}^2\\
	&=\min_{(\dd) \in \spreadset}\abs{\af_{\altaltlogon}\mathopen{}\left(\frac{\delay}{\beta},\beta\doppler\right)}^2\\
	&=\min_{(\dd) \in \altspreadset} \abs{\af_{\altaltlogon}(\dd)}^2.
\end{align*}
Similarly, we have 
\begin{align*}
\maxsumambsquare &= \max_{(\dd) \in \spreadset}  \mathop{\sumdtime\sumdfreq}_{(\dtdf)\neq(0,0)}\abs{\af_{\logon}(\delay -\dtime\tstep,\doppler -\dfreq\fstep)  }^{2} \\
&=\max_{(\dd) \in \spreadset}  \mathop{\sumdtime\sumdfreq}_{(\dtdf)\neq(0,0)}\abs{\af_{\altaltlogon}\mathopen{}\left(\frac{\delay -\dtime\tstep}{\beta},\beta(\doppler -\dfreq\fstep)\right)  }^{2}\\
&=\max_{(\dd) \in \spreadset}  \mathop{\sumdtime\sumdfreq}_{(\dtdf)\neq(0,0)}\abs{\af_{\altaltlogon}\mathopen{}\left(\frac{\delay}{\beta}-\dtime\alttstep,\beta\doppler -\dfreq\altfstep\right)  }^{2}\\
&=\max_{(\dd )\in \altspreadset}  \mathop{\sumdtime\sumdfreq}_{(\dtdf)\neq(0,0)}\abs{\af_{\altaltlogon}\mathopen{}\left(\delay-\dtime\alttstep,\doppler -\dfreq\altfstep\right) }^{2}.
\end{align*}
To conclude, we note that for~$\beta=\sqrt{\tstep/\fstep}$ and under the assumption~$\maxDoppler\tstep=\maxDelay\fstep$, we get~$\alttstep=\altfstep=\sqrt{\tstep\fstep}$, and~$\altmaxDelay=\altmaxDoppler=\sqrt{\spread}/2$, which implies~\eqref{eq:LB_trasformation}.


\begin{IEEEbiographynophoto}{Giuseppe Durisi}(S'02--M'06--SM'12)
received the Laurea degree summa cum laude and the Doctor degree both from Politecnico di Torino, Italy, in 2001 and 2006, respectively. From 2002 to 2006, he was with Istituto Superiore Mario Boella, Torino, Italy. From 2006 to 2010 he was a postdoctoral researcher at ETH Zurich, Zurich, Switzerland. Since 2010 he has been an assistant professor at Chalmers University of Technology, Gothenburg, Sweden. He held visiting researcher positions at IMST (Germany), University of Pisa (Italy),  ETH Zurich (Switzerland), and Vienna University of Technology (Austria). 

Dr. Durisi is a Senior Member of the IEEE, he served as TPC member in several IEEE conferences, and is currently publications editor of the IEEE Transactions on Information Theory. His research interests are in the areas of  information theory, communication  theory, and compressive sensing.   
\end{IEEEbiographynophoto}

\begin{IEEEbiographynophoto}{Veniamin I. Morgenshtern}
	 was born in Leningrad, Russia, in 1982. He studied mathematics and software engineering at Saint-Petersburg State University, Russia, where he received the Dipl. Math. degree with honors in 2004. He then joined the Communication Technology Laboratory at ETH Zurich, Switzerland, as a research assistant. In 2007 he  was a visiting researcher at the University of Illinois at Urbana-Champaign, U.S.A.
	He graduated from ETH Zurich in 2010, receiving the Dr. Sc. degree. From 2010 to 2012, Dr. Morgenshtern was a postdoctoral researcher at ETH Zurich. He is currently a postdoctoral researcher with the Department of Statistics, Stanford University, Stanford, CA. His research interests are in information theory, communication theory, mathematical signal processing, and high-dimensional statistics.                   
	
	Dr. Morgenshtern received the ETH Medal for his dissertation. He currently holds a Swiss National Science Foundation scholarship for advanced researchers.
\end{IEEEbiographynophoto}

\begin{IEEEbiographynophoto}
{Helmut B\"olcskei}
(M'98, SM'02, F'09) was born in M\"odling, Austria on May 29, 1970, and received the Dipl.-Ing. and Dr. techn. degrees in electrical engineering from Vienna University of Technology, Vienna, Austria, in 1994 and 1997, respectively. In 1998 he was with Vienna University of Technology. From 1999 to 2001 he was a postdoctoral researcher in the Information Systems Laboratory, Department of Electrical Engineering, and in the Department of Statistics, Stanford University, Stanford, CA. He was in the founding team of Iospan Wireless Inc., a Silicon Valley-based startup company (acquired by Intel Corporation in 2002) specialized in multiple-input multiple-output (MIMO) wireless systems for high-speed Internet access, and was a co-founder of Celestrius AG, Zurich, Switzerland. From 2001 to 2002 he was an Assistant Professor of Electrical Engineering at the University of Illinois at Urbana-Champaign. He has been with ETH Zurich since 2002, where he is Professor of Electrical Engineering. He was a visiting researcher at Philips Research Laboratories Eindhoven, The Netherlands, ENST Paris, France, and the Heinrich Hertz Institute Berlin, Germany. 

His research interests are in information theory, mathematical signal processing, and applied and computational harmonic analysis.
He received the 2001 IEEE Signal Processing Society Young Author Best Paper Award, the 2006 IEEE Communications Society Leonard G. Abraham Best Paper Award, the 2010 Vodafone Innovations Award, the ETH ``Golden Owl'' Teaching Award, is a Fellow of the IEEE, a EURASIP Fellow, and was an Erwin Schr\"odinger Fellow (1999-2001) of the Austrian National Science Foundation (FWF). He was a plenary speaker at several IEEE conferences and served as an associate editor of the IEEE Transactions on Information Theory, the IEEE Transactions on Signal Processing, the IEEE Transactions on Wireless Communications, and the EURASIP Journal on Applied Signal Processing. He is currently editor-in-chief of the IEEE Transactions on Information Theory and serves on the editorial boards of ``Foundations and Trends in Networking'', ``Foundations and Trends in Communications and Information Theory'' and  the IEEE Signal Processing Magazine. He was TPC co-chair of the 2008 IEEE International Symposium on Information Theory and served on the Board of Governors of the IEEE Information Theory Society.
\end{IEEEbiographynophoto}

\end{document}